\documentclass[12pt, aps, superscriptaddress, onecolumn, tightenlines]{revtex4-2}
\setcitestyle{super}

\usepackage[utf8]{inputenc}
\usepackage{graphicx}
\usepackage{bm}
\usepackage{textcomp,gensymb}
\usepackage{lineno}

\usepackage{prettyref}
\newrefformat{fig}{Fig.~\ref{#1}}
\newrefformat{exfig}{Extended Data Fig.~\ref{#1}}

\usepackage[normalem]{ulem}
\usepackage{color}
\usepackage{units}
\usepackage{times}
\usepackage{amsmath,amssymb}
\usepackage{comment}
\usepackage{hyperref}
\hypersetup{
    colorlinks=true,
    linkcolor=blue,
    filecolor=magenta,
    urlcolor=cyan,
    citecolor=red,
    pdftitle={Imaging inter-valley coherent order in magic-angle twisted trilayer graphene},
}
\usepackage{wasysym}
\usepackage{dsfont}
\usepackage{xcolor}

\definecolor{orange}{rgb}{1,0.5,0}

\newcommand{\ba}{\bm{a}} 

\newcommand{\bq}{\bm{q}}
\newcommand{\bA}{\bm{A}}

\newcommand{\bK}{\bm{K}}
\newcommand{\bG}{\bm{G}}
\newcommand{\bk}{\bm{k}}
\newcommand{\br}{\bm{r}}
\newcommand{\bg}{\bm{g}}
\newcommand{\bR}{\bm{R}}

\newcommand{\bI}{\bm{I}}

\newcommand{\cM}{{\cal M}}
\newcommand{\cE}{{\cal E}}
\newcommand{\cT}{{\cal T}}
\newcommand{\cC}{{\cal C}}
\newcommand{\cK}{{\cal K}}

\usepackage{amsmath,amssymb}
\usepackage{mathtools}

\renewcommand{\figurename}{\textbf{Fig.}}
\newcommand{\beginsupplement}{%
        \setcounter{table}{0}
        \renewcommand{\table}{\arabic{table}|}%
        \renewcommand{\thesection}{\arabic{section}}
        \setcounter{figure}{0}
        \renewcommand{\figurename}{\textbf{Supplementary Information Fig.}}
        
     }

\newcommand{\caltechPH}{Department of Physics, California Institute of Technology, Pasadena, California 91125, USA}
\newcommand{\caltechAPH}{T. J. Watson Laboratory of Applied Physics, California Institute of Technology,
  1200 East California Boulevard, Pasadena, California 91125, USA}
\newcommand{\caltechTH}{Walter Burke Institute for Theoretical Physics, California Institute of Technology, Pasadena, California 91125, USA}
\newcommand{\caltechIQIM}{Institute for Quantum Information and Matter, California Institute of Technology, Pasadena, California 91125, USA}
\newcommand{\UCSB}{Department of Physics, University of California at Santa Barbara, Santa Barbara, California 93106, USA}
\newcommand{\nims}{National Institute for Materials Science, Namiki 1-1, Tsukuba, Ibaraki 305 0044, Japan}

\begin{document}

\title{Imaging inter-valley coherent order in magic-angle twisted trilayer graphene}

\author{Hyunjin Kim}
\thanks{These two authors contributed equally}
\affiliation{\caltechAPH}
\affiliation{\caltechIQIM}
\affiliation{\caltechPH}
\author{Youngjoon Choi}
\thanks{These two authors contributed equally}
\affiliation{\UCSB}
\author{\'Etienne Lantagne-Hurtubise}
\affiliation{\caltechIQIM}
\affiliation{\caltechPH}
\author{Cyprian Lewandowski}
\affiliation{\caltechIQIM}
\affiliation{\caltechPH}
\affiliation{National High Magnetic Field Laboratory, Tallahassee, Florida, 32310, USA}
\affiliation{Department of Physics, Florida State University, Tallahassee, Florida 32306, USA}
\author{Alex Thomson}
\affiliation{\caltechIQIM}
\affiliation{\caltechPH}
\affiliation{\caltechTH}
\affiliation{Department of Physics, University of California, Davis, California 95616, USA}
\author{Lingyuan Kong}
\affiliation{\caltechAPH}
\affiliation{\caltechIQIM}
\author{Haoxin Zhou}
\affiliation{\caltechAPH}
\affiliation{\caltechIQIM}
\author{Eli Baum}
\affiliation{\caltechAPH}
\affiliation{\caltechIQIM}
\author{Yiran Zhang}
\affiliation{\caltechAPH}
\affiliation{\caltechIQIM}
\affiliation{\caltechPH}
\author{Ludwig Holleis}
\affiliation{\UCSB}
\author{Kenji Watanabe}
\affiliation{\nims}
\author{Takashi Taniguchi}
\affiliation{\nims}
\author{Andrea F. Young}
\affiliation{\UCSB}
\author{Jason Alicea}
\affiliation{\caltechIQIM}
\affiliation{\caltechPH}
\affiliation{\caltechTH}
\author{Stevan Nadj-Perge}
\email{Correspondence: hyunjin@caltech.edu; s.nadj-perge@caltech.edu}
\affiliation{\caltechAPH}
\affiliation{\caltechIQIM}

\begin{abstract}
\end{abstract}
\maketitle

\textbf{Magic-angle twisted trilayer graphene (MATTG) exhibits a range of
strongly correlated electronic phases that spontaneously break its underlying 
symmetries\cite{parkTunableStronglyCoupled2021, haoElectricFieldTunable2021}. 
The microscopic nature of these phases and their residual symmetries stands 
as a key outstanding puzzle whose resolution promises to shed light on the origin of 
superconductivity in twisted materials. Here we investigate correlated phases  
of MATTG using scanning tunneling microscopy and identify striking signatures 
of interaction-driven spatial symmetry breaking. In low-strain samples, over 
a filling range of about 2-3 electrons or holes per moir\' e unit cell, 
we observe atomic-scale reconstruction of the graphene lattice that 
accompanies a correlated gap in the tunneling spectrum. This short-scale 
restructuring  appears as a Kekul\'e supercell---implying spontaneous inter-valley 
coherence between electrons---and persists in a wide range of magnetic 
fields and temperatures that coincide with the development of the gap. 
Large-scale maps covering several moir\'e unit cells further reveal a 
slow evolution of the Kekul\'e pattern, indicating that atomic-scale reconstruction 
coexists with translation symmetry breaking at the much longer moir\'e scale. 
We employ auto-correlation and Fourier analyses to extract the intrinsic 
periodicity of these phases and find that they are consistent 
with the theoretically proposed incommensurate Kekul\'e spiral order\cite{kwanKekulSpiralOrder2021, 
wagnerGlobalPhaseDiagram2022}. Moreover, 
we find that the wavelength characterizing moir\'e-scale modulations  
monotonically decreases
with hole doping away from half-filling of the bands 
and
depends only weakly on the magnetic field. Our results provide essential 
insights into the nature of MATTG correlated phases in the presence of strain 
and imply that superconductivity emerges from an inter-valley coherent 
parent state.}

Spontaneous symmetry breaking lies at the foundation of condensed matter 
physics, as the emergence of novel quantum phases often accompanies symmetry 
reduction. Superconductivity and magnetism provide canonical examples 
that emerge when charge conservation and spin-rotation symmetry are respectively broken. 
In the strongly correlated realm, superconductivity commonly occurs in conjunction 
with other forms of symmetry breaking\cite{fradkinColloquiumTheoryIntertwined2015, 
fernandesIronPnictidesChalcogenides2022}, and unraveling their intricate relation presents a profound 
challenge relevant for many platforms---including the growing family of twisted graphene 
multilayers\cite{parkTunableStronglyCoupled2021, haoElectricFieldTunable2021,
zhangPromotionSuperconductivityMagicangle2022, parkRobustSuperconductivityMagicangle2022}.  
Scanning tunneling microscopy (STM) is a well-established tool for identifying certain 
symmetry-broken states, particularly those that leave direct signatures in real space via the local 
density of states distribution. Nevertheless, the inherently difficult task of creating large, 
sufficiently clean and low-strained areas in magic-angle twisted multilayers has so far hindered the 
ability of STM to generate spatial maps of electronic structures sufficient for unambiguously diagnosing 
microscopic symmetry-breaking order. Previous studies have therefore focused mostly on observing 
spectroscopic signatures combined with basic structural 
characterization\cite{kerelskyMaximizedElectronInteractions2019, choiElectronicCorrelationsTwisted2019, 
xieSpectroscopicSignaturesManybody2019, jiangChargeOrderBroken2019}, especially in the context of MATTG
\cite{kimEvidenceUnconventionalSuperconductivity2022, turkelOrderlyDisorderMagicangle2022}. 
Only very recently, this challenge has been resolved in twisted 
bilayers\cite{nuckollsQuantumTexturesManybody2023}, while MATTG has not been 
explored in the context of symmetry-breaking orders.

\begin{figure}[p]
\begin{center}
    \includegraphics[width=16cm]{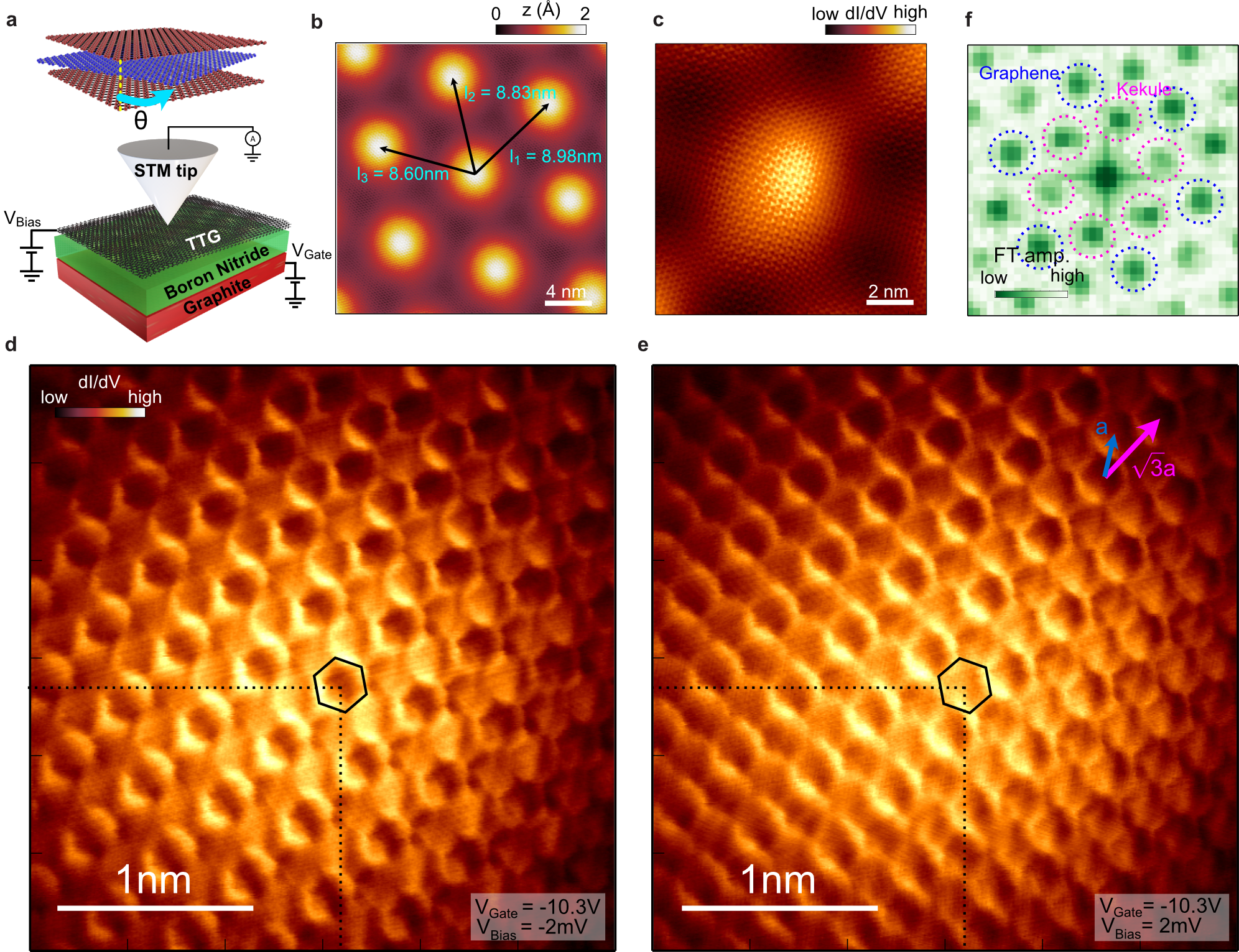}
\end{center}
\caption{ {\bf Overview of the experiment and atomically resolved maps revealing Kekul\'e pattern.} 
  {\bf a}, Schematic depicting twisted trilayer graphene (up) and device geometry used for the STM measurements (down). {\bf b}, Atomically resolved topography showing the moir\' e lattice used to extract hetero-strain magnitude $| \epsilon | \approx 0.12\%$ and twist angle $\theta=1.60\degree$. {\bf c}, Tunneling conductance map taken at $V_{\rm Gate} = -9$V and $V_{\rm Bias} = -2$mV at the AAA site showing atomic resolution. {\bf d}, {\bf e}, $dI/dV$ map measured at fixed $V_{\rm Gate} = -10.3 V$ at negative $V_{\rm Bias} = -2$mV ({\bf d}) and positive $V_{\rm Bias} = 2$mV ({\bf e}) further zoom in into AAA site and showing contrast inversion upon change of ${V_{\rm Bias}}$. {\bf f}, Fourier transformation of \prettyref{fig: fig1}d showing well-resolved peaks corresponding to graphene lattice and Kekul\'e reconstruction. Data in panels {\bf d} and {\bf e} were taken at $T = 400$~mK. Unless otherwise specified, data is taken at $T=2$~K.}
\label{fig: fig1}
\end{figure}

\prettyref{fig: fig1}a sketches the STM measurement setup and the structure of twisted trilayer graphene where 
the second layer is twisted by the angle $\theta$, and the first and third layers are aligned. 
This configuration is known to exhibit an electronic structure that, in the absence of a perpendicular 
electric field, hosts twisted bilayer graphene-like flat bands and an additional set of dispersive Dirac 
cones\cite{khalafMagicAngleHierarchy2019, carrUltraheavyUltrarelativisticDirac2020}. We focus on an area 
with low heterostrain ($\epsilon \approx 0.12\%$) and twist angle 
$\theta = 1.60\degree$, close to the magic-angle value for TTG (see Methods and \prettyref{exfig: Method} and \prettyref{exfig: AFM} for fabrication details). 
High-resolution imaging resolving the atomic scale of TTG near an AAA site---for which carbon atoms 
on all three layers are aligned (\prettyref{fig: fig1}b,c)---reveals a honeycomb structure 
with $0.246$ nm lattice constant accompanied by a more subtle larger-scale atomic modulation. 
This modulation, while visible in topography, is more prominent in the $dI/dV$ spatial maps 
(\prettyref{fig: fig1}d,e) that show a clear lattice 
tripling pattern, the intensity of which depends sensitively on the applied bias voltage 
$V_{\rm Bias}$, with the contrast approximately inverting upon changing the sign of
$V_{\rm Bias}$ (see \prettyref{exfig: Kekule_bias}). Fourier analysis of the $dI/dV$ maps 
further confirms the periodicity of this pattern (\prettyref{fig: fig1}f): a set of six outer peaks 
corresponding to the graphene lattice is accompanied by six inner peaks that correspond to 
$1/\sqrt{3}$ of the graphene reciprocal lattice vectors rotated by $30\degree$. This enlargement of the 
graphene unit cell into a $30\degree$-rotated  $\sqrt{3}\times\sqrt{3}$ supercell corresponds to 
a so-called Kekul\'e distortion previously observed in monolayer graphene decorated with metallic 
adatoms\cite{gutierrezImagingChiralSymmetry2016} or in the quantum Hall 
regime\cite{liuVisualizingBrokenSymmetry2022, coissardImagingTunableQuantum2022}. 
In general, the observation of Kekul\'e distortion implies a reduction of the Brillouin 
zone arising from a coherent scattering of electrons between the $K$ and $K^\prime$ valleys, 
commonly referred to as inter-valley coherence. 

\begin{figure}[p]
\begin{center}
    \includegraphics[width=15cm]{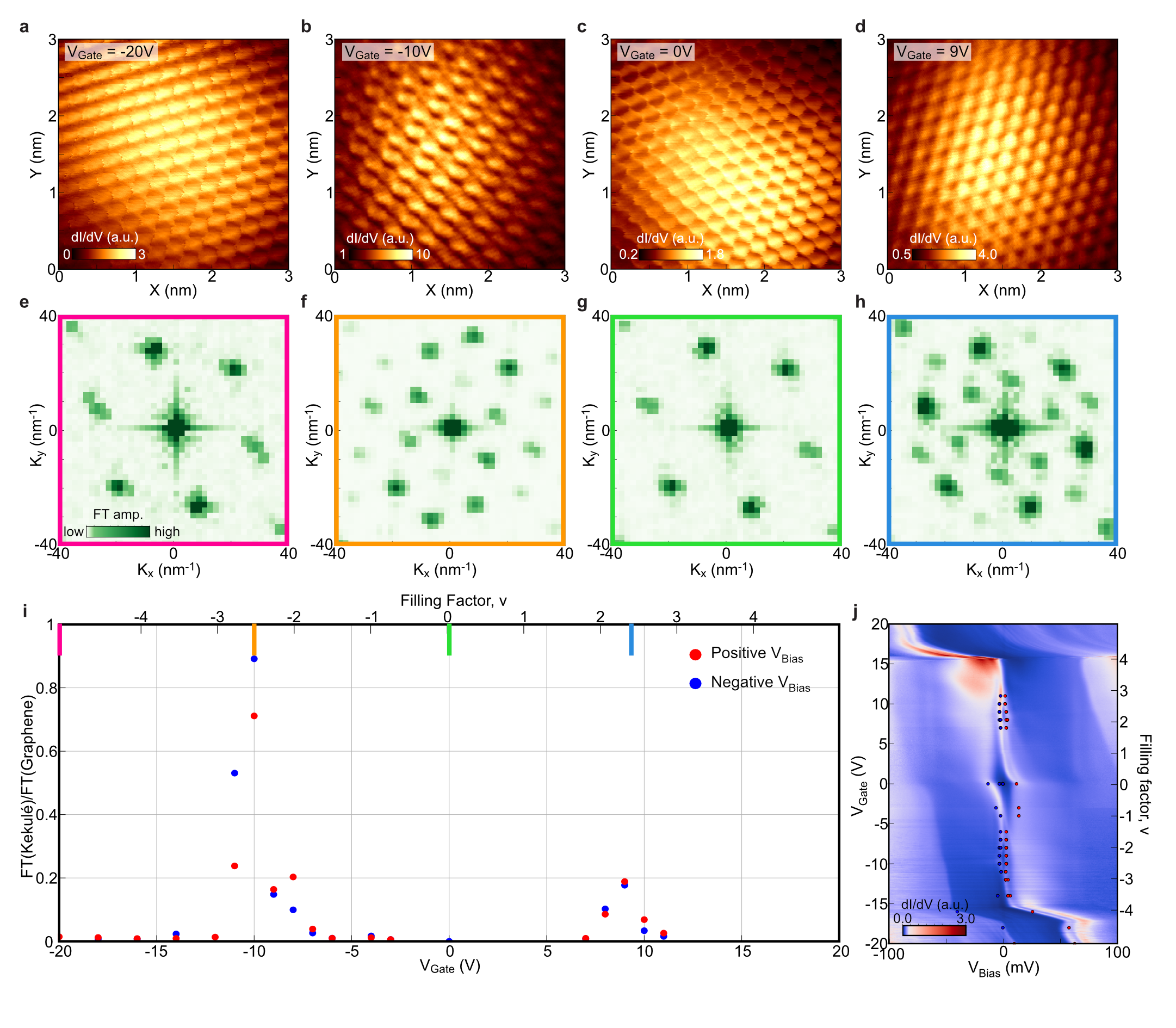}
\end{center}
\caption{
{\bf $\mathbf{V_{\rm Gate}}$ dependent mapping of Kekul\' e 
order on MATTG.} {\bf a-d}, Real space $dI/dV$ map at $V_{\rm Gate} = -20$V
({\bf a}), $-10$V ({\bf b}), $0$V ({\bf c}), $9$V ({\bf d}), taken at 
$V_{\rm Bias} = 63$mV ({\bf a}), $-3$mV ({\bf b}), $-13$mV ({\bf c}), 
$3$mV ({\bf d}), tracking the evolution of the flat band density of states. 
{\bf e-h}, 
Fourier transformation of \prettyref{fig: fig2}a-d zoomed in to show the peaks
corresponding to graphene and Kekul\'e reciprocal lattice vectors. In addition 
to graphene and Kekul\'e peaks, in panels {\bf f} and {\bf h} we also clearly 
resolve higher-order peaks. {\bf i}, Intensity 
of the peaks at Kekul\'e reciprocal lattice vector normalized by the intensity of 
the peaks at graphene reciprocal lattice vector as a function of $V_{\rm Gate}$. The intensities at all six Kekul\'e peaks are summed up and divided by the sum of six graphene peak intensities. Blue and red 
dots correspond to averaged data across positive (negative) 
$V_{\rm Bias}$ values for each $V_{\rm Gate}$ (all the values are marked in \prettyref{fig: fig2}j). 
{\bf j}, $V_{\rm Gate}$ dependent $dI/dV$ spectroscopy measured on the same area where \prettyref{fig: fig2}a-h is measured. 
}
\label{fig: fig2}
\end{figure}

The observed Kekul\'e distortion strongly depends on gate voltage, i.e., filling (\prettyref{fig: fig2}). 
Focusing still on the vicinity of the AAA sites, when the MATTG  Fermi level is in the remote bands 
(\prettyref{fig: fig2}a,e) or near charge neutrality (\prettyref{fig: fig2}c,g), the Kekul\'e distortion 
is completely absent. In contrast, a very prominent distortion arises at filling factors of 
around $-3<\nu<-2$ (\prettyref{fig: fig2}b,f) and $2<\nu<3$ (\prettyref{fig: fig2}d,h). 
\prettyref{fig: fig2}i summarizes the observed relative Kekul\'e peak intensity compared to the graphene 
lattice peaks. Strikingly, the regions of non-zero intensity match well with the regions that host either 
an insulating gap or pseudogap in the gate spectroscopy (\prettyref{fig: fig2}j; see also
\prettyref{exfig: Sc_sample} for data on another sample). The relative 
intensity of the Kekul\'e peaks depends strongly on the bias voltage  
and is approximately maximized when $V_{{\rm Bias}}$ matches the LDOS peak accompanying the (pseudo) gap 
(see \prettyref{exfig: Kekule_bias}). Moreover, at higher temperatures, as the gap is suppressed, the 
visibility of Kekul\' e peaks also diminishes (\prettyref{exfig: Kekule_T}). Strong dependence 
of the reconstruction on $V_{\rm Gate}$, $V_{\rm Bias}$ and the temperature indicate  
its electronic origin and rule out the possibility that the observed Kekul\' e peaks arise from 
impurities or structural deformations. 
 
Various inter-valley coherent (IVC) phases have been theoretically proposed as 
the ground state of magic-angle twisted bilayers\cite{kangStrongCouplingPhases2019, 
bultinckGroundStateHidden2020, lianTwistedBilayerGraphene2021, kwanKekulSpiralOrder2021}, 
with a limited number of theoretical calculations specific to 
MATTG\cite{christosCorrelatedInsulatorsSemimetals2022}. We consider the family of 
such candidate IVC phases, motivated by Ref.~\citenum{hongDetectingSymmetryBreaking2022, calugaruSpectroscopyTwistedBilayer2022}
and the close band-structure and spectroscopic resemblances between twisted bilayers 
and trilayers. 
One member of this family---dubbed the Kramers IVC phase---was predicted to appear in unstrained 
samples~\cite{kangStrongCouplingPhases2019, bultinckGroundStateHidden2020, lianTwistedBilayerGraphene2021} 
and manifests as a magnetization density wave that generates Kekul\'e peaks in the LDOS only at finite 
magnetic fields~\cite{hongDetectingSymmetryBreaking2022, calugaruSpectroscopyTwistedBilayer2022}. 
Since we observe pronounced Kekul\'e 
peaks already at zero magnetic field, this IVC order can be ruled out in our MATTG samples.  Two other 
leading candidates include the time-reversal-symmetric IVC (T-IVC) and incommensurate Kekul\'e spiral 
(IKS) phases; both produce charge density waves that generate LDOS Kekul\'e peaks at zero magnetic field, 
compatible with our measurements. 
 
\begin{figure}[p]
\begin{center}
    \includegraphics[width=15cm]{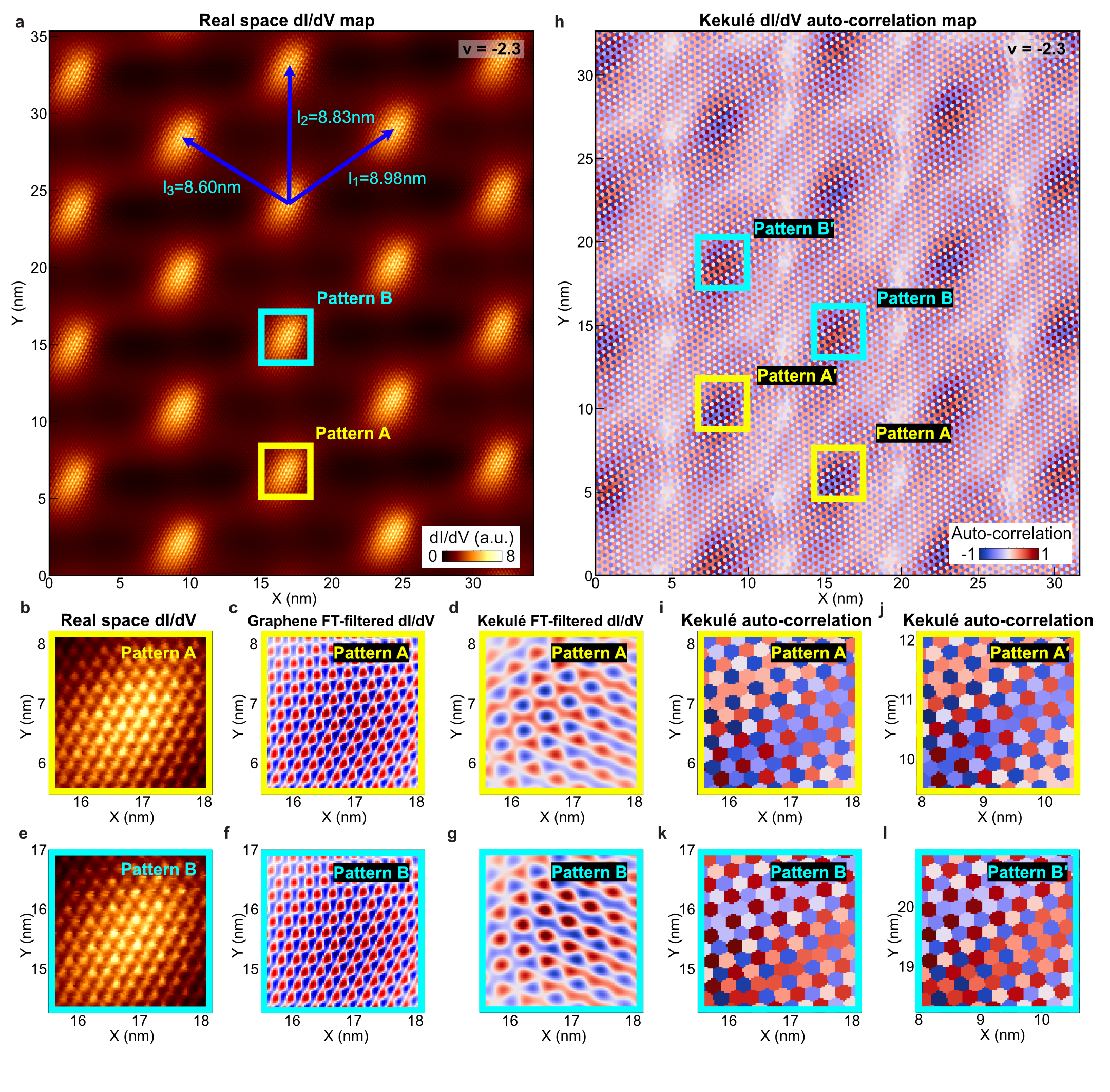}
\end{center}
\caption{{\bf Evidence of moir\' e translation symmetry breaking.} 
{\bf a}, Large-scale area $dI/dV$ scan around $\nu = -2.3$. 
{\bf b,e}, $2.2$ nm $\times$ $2.2$ nm window from \prettyref{fig: fig3}a centered around one AAA site ({\bf b}) and its nearest neighbor AAA site along $l_{2}$ direction ({\bf e}). 
{\bf c,f}, Corresponding graphene FT filtered $dI/dV$ map for \prettyref{fig: fig3}b,e shows the precise alignment of the atomically resolved signals.
{\bf d,g}, Kekul\'e FT filtered $dI/dV$ map  for \prettyref{fig: fig3}b,e exhibiting inverted contrast between neighboring AAA sites.
{\bf h}, Kekul\'e $dI/dV$ auto-correlation map of \prettyref{fig: fig3}a.
{\bf i,j}, $2.2$ nm by $2.2$ nm window taken from Kekul\'e auto-correlation map centered around AAA site showing pattern A ({\bf i}) and neighboring AAA site along $l_{3}$ direction showing almost similar pattern ({\bf j}).
{\bf k,l}, $2.2$ nm by $2.2$ nm window taken from Kekul\'e auto-correlation map centered around AAA site showing pattern B ({\bf k}) and neighboring AAA site along $l_{3}$ direction showing almost similar pattern ({\bf l}).
}
\label{fig: fig3}
\end{figure}

To establish more precisely the nature of the correlated state, we investigate variations of the Kekul\'e pattern across neighboring moir\' e unit cells by taking 
large-area $dI/dV$  maps (taken at $\nu=-2.3$) covering a $36$ nm $\times$ $36$ nm region that 
encompasses twenty moir\' e AAA sites (\prettyref{fig: fig3}a). Focusing first on 
two regions around neighboring AAA sites (\prettyref{fig: fig3}b,e, marked by 
a yellow and blue square in \prettyref{fig: fig3}a), we decompose the map using 
Fourier transform (FT) based filtering to separate the spatial evolution of the 
Kekul\'e distortions from the underlying graphene lattice; 
see methods for details. These regions are chosen to have graphene lattices precisely 
atomically aligned (\prettyref{fig: fig3}c,f). Under these conditions, while the raw data show only 
subtle differences, FT-filtered Kekul\'e patterns clearly show that bright spots (high intensity) 
in the yellow window (\prettyref{fig: fig3}d) become dark spots (low intensity) in the blue window (\prettyref{fig: fig3}g) 
and vice-versa. This dichotomy establishes that the Kekul\'e distortion between two neighboring 
moir\'e sites changes significantly.

To quantify the spatial variation across the entire mapped region, 
we extend the above method and create Kekul\'e auto-correlation maps (\prettyref{fig: fig3}h).
Here, we fix a small region of approximately $1$ nm $\times$ $1$ nm and 
separately auto-correlate graphene 
lattice- and Kekul\'e-filtered periodicities with a `moving window' of the same size that 
samples different map regions. The auto-correlation is normalized such that perfectly correlated 
regions correspond to $1$ while perfectly anti-correlated regions correspond to $-1$. 
We assign Kekul\'e auto-correlation values to the nearest points at which atomic auto-correlation
exceeds a threshold of one-half (the conclusions do not depend on the exact threshold value). 
This procedure ensures alignment of the underlying graphene lattices when comparing Kekul\'e 
modulation across different parts of the map. The zoom-in to the \prettyref{fig: fig3}h map 
reveals a rapid evolution of the auto-correlation and the atomic scale lattice tripling 
(3-unit cell) pattern  (\prettyref{fig: fig3}i,j,k,l; see methods for further discussion). 
On the moir\'e length scale, this Kekul\'e map shows clear stripe-like red-blue patterns along 
the $l_1$ and $l_2$ directions but shows weak dependence along the $l_3$ direction. 
The observed periodicity reflects modulation at a wavevector $\bq_{\rm Kekul\Acute{e}}$ that is perpendicular to 
the $l_3$ direction with a magnitude that is approximately half of a moir\' e 
reciprocal lattice vector (for this filling factor)---corresponding to a near doubling of the moir\' e unit cell.

\begin{figure}[p]
\begin{center}
    \includegraphics[width=15cm]{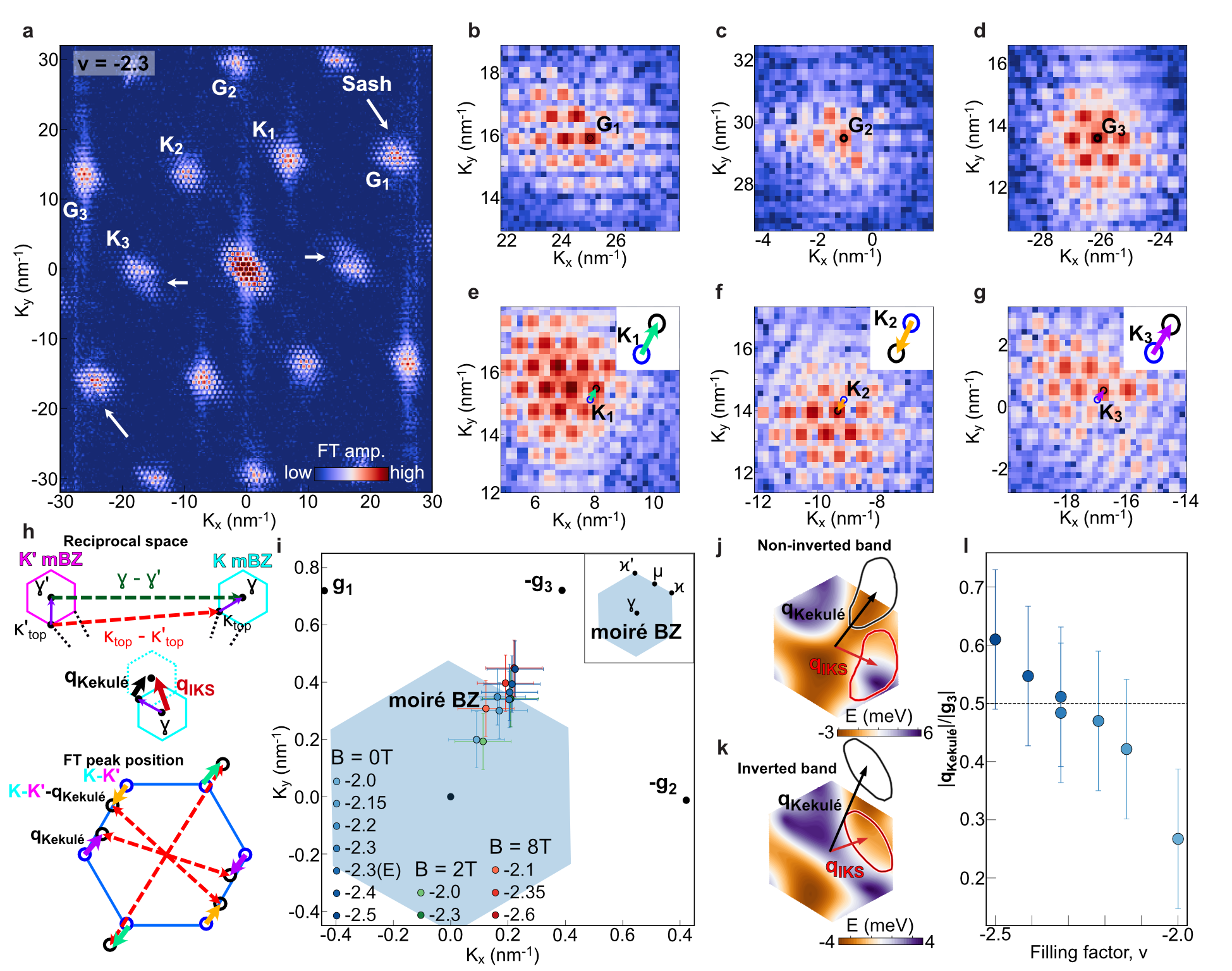}
\end{center}
\caption{{\bf Incommensurate Kekul\'e spiral wavevector extracted from the Fourier transformation maps}. 
{\bf a}, Fourier transformation of the real space $dI/dV$ map at $\nu = -2.3$ and $V_{\rm Bias} = -2$mV. Fourier transform peaks at graphene reciprocal lattice vector site is decorated with moir\'e satellite peaks are named $G_1$, $G_2$, $G_3$, while peaks around Kekul\'e reciprocal lattice vector position are named $K_1$, $K_2$, $K_3$.
{\bf b-d}, Zoom-in images of \prettyref{fig: fig4}a where exact graphene reciprocal lattice vector positions are marked as black dots. 
{\bf e-g}, Zoom-in images of \prettyref{fig: fig4}a around $K_1$, $K_2$, $K_3$, where expected Kekul\'e reciprocal lattice vector positions are marked as black circles. The nearest satelite peaks that are used to extract the modulation wavevector $\bq_{\rm Kekul\acute{e}}$ are marked as blue circles. Arrows denote the direction of $\bq_{\rm Kekul\acute{e}}$ for each $K_1$, $K_2$, $K_3$.
{\bf h}, Schematics depicting the hybridization of $\bK$ and $\bK^{\prime}$ moir\'e BZ for two different cases (upper panel). $\gamma$ to $\gamma^{\prime}$ (green vector) gives Kekul\'e pattern that is commensurate with the moir\'e lattice while $\kappa_{top}$ to $\kappa_{top}^{\prime}$ (red vector) gives spatially uniform Kekul\'e pattern on the graphene scale. Middle panel shows the relation between $\bq_{\rm IKS}$ and $\bq_{\rm Kekul\acute{e}}$. Lower panel shows schematics of FT peaks corresponding to inter-valley coherent state with finite Kekul\'e modulation $\bq_{\rm Kekul\acute{e}}$.
{\bf i}, Extracted modulation wavevector $\bq_{\rm Kekul\acute{e}}$ as a function of both $V_{\rm Gate}$ and out-of-plane magnetic field of $2$T and $8$T. Black dots denote the positions of moir\' e reciprocal lattice vectors $g_1$, $g_2$, $g_3$, which are perpendicular to $l_1$, $l_2$, $l_3$ (respectively) in \prettyref{fig: fig1}b. Blue hexagon is the moir\'e Brillouin zone (mBZ) calculated with the $g_1$, $g_2$, $g_3$ extracted from the experiment. 
{\bf j} and {\bf k}, Calculated band structure of the top-most valence-band in valley $\bK$ in the regime of small Hartree correction ($\epsilon_r=30$, panel {\bf j}), where the flat band is non-inverted, and in the regime of large Hartree correction ($\epsilon_r = 15$, panel {\bf k}), where the flat band is inverted around the $\gamma$ point of the mBZ. The black and red arrows respectively indicate the theoretically optimal $\bq_{\rm Kekul\acute{e}}$ and $\bq_{\rm IKS}$ wavevectors, with contours showing the corresponding confidence interval (see SM Section ~\ref{app:strained_bands} for details.)
{\bf l}, Size of $\bq_{\rm Kekul\acute{e}}$  normalized by the size of the moir\' e reciprocal lattice vector $g_3$ which is 
almost aligned with $\bq_{\rm Kekul\acute{e}}$. Black dashed line displays the mBZ boundary.}
\label{fig: fig4}
\end{figure}

While the auto-correlation analysis shown above demonstrates that translation symmetry 
on the moir\'e length scales is broken, obtaining the modulation periodicity directly 
from real space maps is challenging when the order is far from commensurate. We, therefore, 
turn to a Fourier-space analysis to extract the modulation wavevectors arising over a broad 
filling range. \prettyref{fig: fig4}a shows the Fourier transform of the map shown in \prettyref{fig: fig3}a. 
In contrast to the FT images of the small real-space areas, here the graphene reciprocal lattice and Kekul\'e peaks are decorated by satellite peaks---reflecting the longer wavelength moir\' e pattern in the original image.
Identifying the exact graphene lattice vector peak among the cluster of satellite 
peaks is non-trivial, as even small errors in the STM calibration or possible homostrain 
effects within the top graphene layer can alter it. We overcome this problem by examining the higher-order reciprocal 
lattice vectors, which exhibit smaller clusters in Fourier space and thus allow for more precise 
extraction (see \prettyref{exfig: Higher_order_peaks}). In \prettyref{fig: fig4}b-d, 
we mark the exact position of the graphene reciprocal lattice vectors $\bG_{1,2,3}$. %
Next, based on these vectors, we can accurately estimate the positions of the FT peaks that would arise for a uniform Kekule 
pattern on the graphene scale as 
$\bK_1 = (\bG_{1}+\bG_{2})/3$, $\bK_2 = (\bG_{2}+\bG_{3})/3$, and $\bK_3 = (\bG_{3}-\bG_{1})/3$ 
(blue circles in \prettyref{fig: fig4}e-g). None of the satellite peaks in the FT signal line up with $\bK_{1,2,3}$, 
though crucially, the observed displacement of all three sets of peaks relative to $\bK_{1,2,3}$ can 
be accounted for by a \emph{single} wavevector  $\bq_{\rm Kekul\acute{e}}$.

The extracted Kekul\'e modulation wavevector $\bq_{\rm Kekul\Acute{e}}$  does not change with $V_{\rm Bias}$ but evolves 
monotonically with $V_{\rm Gate}$, and is generally incommensurate with the moir\'e potential. Remarkably, upon hole doping from $\nu = -2$ to $-2.5$, 
$\bq_{\rm Kekul\Acute{e}}$ always points along the $\bg_{3}$ moir\' e reciprocal lattice vector within 
error bars (\prettyref{fig: fig4}i) but increases in magnitude, eventually crossing the moir\'e
Brillouin zone boundary near the commensurate point $\bg_3/2$ (\prettyref{fig: fig4}l).  
Note that these results are consistent with the real-space auto-correlation analysis 
in \prettyref{fig: fig3}. Furthermore, the observed modulation is quite robust to magnetic fields: 
By performing similar mappings in moderate ($2$T) and high ($8$T) out-of-plane magnetic 
fields we find that spatially varying Kekul\'e distortions are still present (\prettyref{exfig: Kekule_B})
and that $\bq_{\rm Kekul\Acute{e}}$ only slightly shifts its direction from its zero field value. Finally, our procedure gives consistent results for $\bq_{\rm Kekul\Acute{e}}$ for maps 
with different sizes and taken at various $V_{\rm Bias}$.

We now discuss the implications of our observations for the nature of the correlated state between $\nu = -2$ and $-2.5$. The identification of a Kekul\'e pattern modulating slowly with a generally incommensurate, doping-dependent wavevector suggests the presence of IKS order.
Theoretically, the IKS state arises out of an inter-valley nesting instability at a single wavevector $\bq_{\rm IKS}$ 
in the presence of small heterostrain ($\sim0.1-0.2\%$)~\cite{kwanKekulSpiralOrder2021, wagnerGlobalPhaseDiagram2022, wangKekulSpiralOrder2022}---comparable to the values of our MATTG samples.
The resulting order yields a lattice-tripling pattern that slowly varies between moir\' e lattice sites with wavelength $2\pi/|\bm{q}_{\rm IKS}|$ along a direction set by $\bq_{\rm IKS}$ (see Fig.~\ref{fig: fig4}h and SM section \ref{app:LDOS_results} for more discussion and theoretical modeling).
Importantly, this vector is distinct from the extracted modulation wavevector $\bq_{\rm Kekul\acute{e}}$: whereas $\bq_{\rm IKS}$ characterizes variations relative to the moiré lattice, $\bq_{\rm Kekul\acute{e}}$ instead measures the modulation of the lattice-tripling order relative to the graphene lattice. 
The two vectors are
related by a momentum shift that connects the mini-BZ center $\gamma$ and the mini-BZ corners $\kappa$ (see schematic in Fig.~\ref{fig: fig4} h).

Because IKS order arises from an inter-valley nesting instability, it is greatly influenced by the structure of the flat bands~\cite{kwanKekulSpiralOrder2021} ---especially the location of 
the flat-band maxima and minima at each valley. In the continuum
model~\cite{Lopes2007, bistritzerMoireBandsTwisted2011}, the valence 
flat-band minima and maxima respectively occur at the moir\'e Brillouin zone $\gamma$ and $\kappa,\kappa'$ points. 
Upon inclusion of heterostrain, such features are distorted in a non-universal fashion that depends on the strain 
angle and magnitude. Electronic interactions, which significantly alter the shape of the flat bands, are 
also crucial to determine the preferred IKS instability. 
Charge inhomogeneity, characterized by the self-consistent Hartree term, tends to invert flat bands around $\gamma$ point\cite{guineaElectrostaticEffectsBand2018, ceaElectronicBandStructure2019, ceaBandStructureInsulating2020,
rademakerChargeSmootheningBand2019, goodwinHartreeTheoryCalculations2020, xieWeakFieldHallResistivity2021}. 
Good agreement with the experiment is found when interaction-induced band inversion is not pronounced. In this case, theoretically 
extracted $\bq_{\rm Kekul\acute{e}}$ that includes measured heterostrain, matches well with the observations within the error bars of both the experimental 
extraction and the theoretical procedure (see SM Sec.~\ref{app:strained_bands} for details).
The calculated modulation wavevector $\bq_{\rm Kekul\acute{e}}$ evolves with hole doping away from $\nu = -2$ 
following experimentally observed trends, with the evolution being more pronounced in the non-inverted regime. 
Finally, we note that the measured pattern in 
\prettyref{fig: fig3} appears very close to a \emph{commensurate} modulation; 
see SM Sec.~\ref{app:comensurate} for a possible mechanism favoring a lock-in of the IKS modulation 
to a nearby commensurate wavevector.

Our experiments reveal several other signatures directly related to symmetry breaking in MATTG. In 
many of the FT maps, we observed suppression of the FT satellite peaks along narrow striped
regions, forming a `sash-like' feature (see white arrows in \prettyref{fig: fig4}a) discussed theoretically in 
Ref.~\citenum{hongDetectingSymmetryBreaking2022} in the context of $C_3$ symmetry breaking (see also \prettyref{exfig: sash_features}). These sashes are resolved at various filling factors
both within the graphene reciprocal lattice vector peaks and Kekul\'e peaks; moreover, consistent 
with $C_3$ symmetry breaking on the graphene lattice scale, they do not appear in all directions of the FT maps. 
This observation is compatible with the IKS ground state around $\nu = -2$.  Interestingly, similar features 
are observed near charge neutrality for $V_{\rm Gate} = 0$~V, where we did not detect lattice tripling order. 
For this gate voltage, we further observed a preferred directionality in graphene bonds and a spatial evolution of the $dI/dV$ signal 
near the AAA site that is consistent with a nematic semi-metallic ground state~\cite{hongDetectingSymmetryBreaking2022} 
(\prettyref{exfig: nsm}). These findings further highlight the ability of STM to distinguish various symmetry-broken 
ground states in moir\' e heterostructures. 

Finally, our spectroscopic detection and characterization of IVC 
order provide fresh insight into the puzzle surrounding the origin of 
superconductivity and the accompanying pseudogap phase in MATTG.  The emergence of 
intervalley coherence at fillings where we previously reported unconventional 
superconductivity\cite{kimEvidenceUnconventionalSuperconductivity2022} 
(see \prettyref{exfig: Sc_sample}) sharply constrains theoretical scenarios.  
Further constraints follow from the surprising observation that the strength of IVC 
order, quantified by the normalized Kekul\' e peak intensity, is maximal in the pseudogap 
regime (around $\nu \approx -2.3$ to $-2.5$) instead of the $\nu = -2$ insulating phase. 
Other enigmatic properties of MATTG have also been observed in this filling range, including the
evolution from U- to V-shaped tunneling spectra observed 
spectroscopically\cite{kimEvidenceUnconventionalSuperconductivity2022}, as well as 
a sharp change in the Ginzburg-Landau coherence length and maximal critical 
temperature observed in transport\cite{parkTunableStronglyCoupled2021, haoElectricFieldTunable2021}. 
Linking these diverse aspects of MATTG phenomenology poses an outstanding challenge for future 
theory and experiments.

\vspace{5pt}
\noindent {\bf Methods:}

\vspace{5pt}

\noindent {\bf Device fabrication:} To fabricate the graphite/hBN/TTG device, the layers are first picked up sequentially using a poly(bisphenol A carbonate) (PC) film on a polydimethylsiloxane (PDMS) block via the typical dry transfer technique at temperatures between 60-100$\degree C$.  The stack then needs to be `flipped' onto a substrate to expose the TTG and electrically contacted it without introducing polymer residues. 

Stack flipping is accomplished using a gold-coated PDMS slide, as shown in \prettyref{exfig: Method}a-f. In this step, the PC film initially supporting the stack is peeled off of the PDMS slide and transferred stack-side down onto PDMS block coated with Ti/Au (3/12nm) via e-beam evaporation. The PC film is then removed using N-methyl-2-pyrrolidone (NMP). The gold surface and the stack are naturally sealed together, preventing the NMP from permeating the interface containing the TTG. Moreover, the evaporated metals prevent any residues from contacting the sample surface.  Because the stiction between gold and the van der Waals stack is weaker that between the van der Waals stack and silica, the stack can then be `dropped’ onto an oxide substrate without the need for further solvent use. 

To apply bias and gate voltages to the sample without contaminating the surface, we use a ‘gold stamping’ technique illustrated in \prettyref{exfig: Method}g-k. First, a SU8 (SU-8 2005, Microchem) photoresist mold is defined on a silicon oxide substrate. Then, PDMS (SYLGARD 184, 10:1) is poured onto the mold and dried, resulting in a patterned stamp after peeling it off from the mold. Gold is evaporated (e-beam evaporation, $10-20nm$) onto the PDMS stamp, which is then pressed down onto the desired area of the sample at 130$\degree C$. As only the highest parts of the stamp make contact with the sample---leaving gold behind--- the rest of the sample surface remains uncontaminated.

Samples B and C were prepared using our previously developed ‘PDMS assisted flipping’ technique \cite{choiElectronicCorrelationsTwisted2019}. The hBN/MATTG stack is picked up using a PC film, peeled off, and flipped onto an intermediate PDMS block. The PC film is then washed away with NMP/isopropyl alcohol and kept under vacuum for several days. The stack is subsequently transferred onto a graphite gate that has been previously dropped on a substrate. Lastly, a graphite contact is exfoliated on PDMS and dropped to connect MATTG to a pre-patterned electrode. 

To compare the quality of sample A to B, we used AFM in AC tapping mode (\prettyref{exfig: AFM}). Sample B exhibited residues on the surface, which became apparent after scanning (cleaning) an area with contact mode due to the appearance of square residue boundaries. However, no signs of residues were observed on sample A before or after scanning the area with contact mode. Once a sample was fabricated, no further annealing was conducted inside the STM chambers. 

\noindent {\bf Conductive AFM Characterization at Room Temperature:}
Conductive Atomic Force Microscopy (cAFM) is a powerful imaging technique used for visualizing moir\' e patterns in twisted heterostructures, as reported in various studies\cite{zhangAbnormalConductivityLowangle2020, huangImagingDualMoireLattices2021}. We employed cAFM to characterize the twist angles and cleanliness of a sample at room temperature prior to cooling it down for STM measurements. A commercial AFM (Asylum Research Cypher) equipped with a conducting tip (ASYELEC-01-R2 from Asylum Research, with a spring constant of approximately 2.8N/m) is utilized. During cAFM imaging, bias voltage of 100mV was typically applied between the tip and the twisted graphene, while the gate was left floating. The typical magic angle twisted trilayer graphene sample exhibits moir\' e pattern surrounded by supermoir\' e stripes, as previously reported in the literature \cite{kimEvidenceUnconventionalSuperconductivity2022} (\prettyref{exfig: AFM}c,f). The region with a moir\' e wavelength of approximately 9 nm, which corresponds to the magic angle of twisted trilayer graphene, was the target during the navigation of the STM tip to the sample. 

\noindent {\bf STM measurements:}
The STM measurements were performed in a Unisoku USM 1300J STM/AFM system using
a Platinum/Iridium (Pt/Ir) tip as in our previous works on
bilayers\cite{choiElectronicCorrelationsTwisted2019,
  choiCorrelationdrivenTopologicalPhases2021,
  choiInteractiondrivenBandFlattening2021}. All reported features are observed
with many (usually at least ten) different microtips. Unless specified otherwise, data 
was taken at temperature $T=2$~K and the
parameters for $dI/dV$ spectroscopy measurements were $V_{\rm Bias} = 100$~mV
and $I = 1$~nA, and the lock-in parameters were modulation voltage
$V_{\rm mod} = 0.2-1$~mV and frequency $f = 973$~Hz. Real space $dI/dV$ maps are 
taken with the constant height mode (feedback turned off, tilt corrected).
The piezo scanner is
calibrated on a Pb(110) crystal and verified by
measuring the distance between carbon atoms. The twist-angle uncertainty is
approximately $\pm0.01\degree$, and is determined by measuring moir\'e wavelengths
from topography. Filling factor assignment has been performed by taking Landau
fan diagrams as discussed previously\cite{choiCorrelationdrivenTopologicalPhases2021} 
or by identifying features corresponding to full-filling and CNP 
LDOS suppression\cite{choiElectronicCorrelationsTwisted2019} in data sets where 
magnetic field dependence was not studied. The deviations between the two methods 
in assigning filling factors are typically within 5\%.

\noindent {\bf Heterostrain extraction:}
The presence of heterostrain deforms the moir\'e lattice from an ideal equilateral triangular lattice.
By experimentally measuring the two moir\'e lattice vectors $\bg_1$ and $\bg_2$, we can extract the magnitude and direction of the heterostrain that best matches the observed moir\'e lattice geometry~\cite{kerelskyMaximizedElectronInteractions2019} (see also SM section \ref{app:heterostrain} for details of strain modeling). The magnitude of the heterostrain is uniquely determined from the moir\'e lattice constants $l_1$, $l_2$ and $l_3$ shown in Fig.~\ref{fig: fig1}b---we obtain a magic angle of $1.60^\circ \pm 0.01^\circ$ and strain magnitude $|\epsilon| = \left( 0.12 \pm 0.04 \right) \% $. The extraction of the strain direction is additionally sensitive to the twist-angle direction (clock-wise vs counter-clockwise), as well as which layer experiences compressive strain. In our experiment, we stacked twisted trilayer graphene such that the middle layer is rotated counter-clockwise with respect to the top and bottom layers. We then extract the direction of the moir\'e lattice vectors $\bg_1$ and $\bg_2$ by comparing to an armchair direction of graphene, or equivalently to the $\bG_1$ vector in the FT. This information allows us to conclude that the middle layer experiences compressive strain (corresponding to negative $\epsilon$ in our conventions). With these additional pieces of information, the strain magnitude $\epsilon = - \left(0.12 \pm 0.04 \right)\%$ strain angle $\varphi = \left( 87 \pm 10 \right)^\circ$ are extracted.

\noindent {\bf Kekul\'e auto-correlation analysis:}
Variation of the Kekul\'e pattern on the neighboring AAA sites is quantified by calculating the correlation between two small images taken at two spatially different parts of the large $dI/dV$ map. 
The correlation, which we define as an element-wise multiplication between two same-sized images, is normalized such that it gives $1$ for two identical images and $-1$ for fully reversed image. 
A 2D auto-correlation map plots the correlation between a fixed window, centered at a particular AAA site, and a moving window that spans the whole $dI/dV$ map. 
The size of the window we choose to calculate the auto-correlation map in \prettyref{fig: fig3}h is $1.2$ nm; the result is insensitive to the size of the window within $0.8$ nm to $1.7$ nm.

A Kekul\'e $dI/dV$ auto-correlation map calculates the correlation in a FT-filtered $dI/dV$ map that only contains signals that are periodic with the Kekul\'e wavevector. 
FT filtering is performed by first identifying six peaks correponding to Kekul\'e distortion ($|\bK_{i}| = \frac{4 \pi}{3 \sqrt{3} a_0}$) and 6 peaks for graphene reciprocal lattice vectors ($|\bG_{i}| = \frac{4 \pi}{3 a_0}$).
An FT-filtered Kekul\'e $dI/dV$ map is produced by filtering out all except the 6 Kekul\'e peaks and then performing an inverse FT, while six graphene reciprocal lattice peaks are used for FT-filtered graphene $dI/dV$ map.
We calculate auto-correlation on the FT-filtered Kekul\'e $dI/dV$ maps and plot only those peaks that are at local maxima in the graphene $dI/dV$ auto-correlation map to trace the evolution of auto-correlation only when the two windows are atomically aligned.

\noindent {\bf Extraction of IKS wavevector:}
Because the IKS ground state is constructed by hybridizing the $K$ and $K^\prime$ valleys with a momentum mismatch $\bq_{\rm IKS}$, FT peaks appear at $\gamma-\gamma^{\prime}-\bq_{\rm IKS}$ or equivalently at $\kappa_{top}-\kappa^{\prime}_{top}-\bq_{\rm Kekul\acute{e}}$, rather than $\kappa_{top}-\kappa^{\prime}_{top}$ for uniform Kekul\'e distortion (see \prettyref{fig: fig4}h).
We extract the IKS wavevector $\bq_{\rm IKS}$ or equivalently $\bq_{\rm Kekul\acute{e}}$ by measuring how much the FT peaks are offset from $\kappa_{top}-\kappa^{\prime}_{top}$.
While in theory, the vector $\kappa_{top}-\kappa^{\prime}_{top}$ can be easily deduced from graphene reciprocal lattice vector peaks in the FT, small uncertainty in piezo calibration and the effect of strain makes it hard to distinguish the central graphene reciprocal lattice vector peak among many satellite peaks with similar intensities.
Although neighboring satellite peaks at $\bG_{i}$ are separated by one moir\'e reciprocal lattice vector $\bg$, their high-order counterparts at $N\bG_{i}$ are apart by $N\bg$.
If our choice of exact $\bG_{i}$ is off by $\bg$, the exact $3\bG_{i}$ will be located $3\bg$ away from the current $\bG_{i}$ (marked as the yellow circle in the \prettyref{exfig: Higher_order_peaks}), which is unlikely considering that the prominent satellite peaks locate within $\pm 2\bg$ from the center of the cluster.
Our choice of $\bG_{1}$ and $\bG_{2}$ in \prettyref{fig: fig4}b,c is the only choice that yields high-order FT peaks at $m\bG_{1}+n\bG_{2}$ ($m$, $n$ are integers from $-3$ to $+3$);
$\bG_{3}$ is automatically determined by $\bG_{3} = \bG_{2}-\bG_{1}$ because we only have two independent reciprocal lattice vectors.
The plotted $\bG_{3}$ (black circle in \prettyref{fig: fig4}d) is precisely on top of an FT peak, consistent with the fact that we do not observe other long-wavelength patterns in FT-filtered `graphene' $dI/dV$ maps.
Once we determine which FT peak corresponds to the exact $\bG_{i}$, the naively expected Kekul\'e FT peak positions (which would occur for T-IVC order) are calculated as $\bK_1 = (\bG_{1}+\bG_{2})/3$, $\bK_2 = (\bG_{2}+\bG_{3})/3$, and $\bK_3 = (\bG_{3}-\bG_{1})/3$.
Any FT peak can be chosen to calculate the $\bq_{\rm Kekul\acute{e}}$, defined as the momentum separation from the calculated $\bK_i$, modulo a moir\'e reciprocal lattice vector $\bg$.
A particular FT peak is chosen to calculate $\bq_{\rm Kekul\acute{e}}$, so that the norm of $\bq_{\rm Kekul\acute{e}}$ does not exceed the norm of $\bg$ and the gate dependence of $\bq_{\rm Kekul\acute{e}}$ can be traced continuously.
Also, three $\bq_{\rm Kekul\acute{e}}$ can be extracted separately from $\bK_{1}$, $\bK_{2}$, and $\bK_{3}$, and all three values match within the experimental range of error.


\noindent {\bf Acknowledgments:}  We thank Nick Bultinck, Sid Parameswaran, Abhay Pasupathy, Ashvin Vishwanath, Senthil Todadri,
and Ali Yazdani for fruitful discussions. We are grateful in particular to Tomohiro Soejima and Mike Zaletel for pointing out subtleties regarding the extraction of $\bq_{\rm IKS}$ through Fourier analysis. {\bf Funding:} This work has been primarily
supported by National Science Foundation (grant no. DMR-2005129);  
Office of Naval Research (grant no. N142112635); and Army Research Office under Grant
Award W911NF17-1-0323. S.N-P. acknowledges support from the Sloan Foundation. 
J.A. and S.N.-P. also acknowledge support of the Institute for
Quantum Information and Matter, an NSF Physics Frontiers Center with support of
the Gordon and Betty Moore Foundation through Grant GBMF1250; \'E. L.-H. and C.L. acknowledge
support from the Gordon and Betty Moore Foundation’s EPiQS Initiative, Grant
GBMF8682 (at Caltech). C.L. acknowledges start-up funds from Florida State 
University and the National High Magnetic Field Laboratory. The National High 
Magnetic Field Laboratory is supported by the National Science Foundation 
through NSF/DMR-1644779 and the state of Florida.  A.T. and J.A. are grateful 
for the support of the Walter 
Burke Institute for Theoretical Physics at Caltech. H.K. acknowledges support
from the Kwanjeong fellowship. L.K. acknowledges support from an IQIM-AWS 
Quantum postdoctoral fellowship. Work at UCSB was supported by the U.S. Department 
of Energy (Award No. DE-SC0020305) and by the Gordon and Betty Moore Foundation under 
award GBMF9471.  This work used facilities supported via the UC Santa Barbara NSF Quantum 
Foundry funded via the Q-AMASE-i program under award DMR-1906325.

\noindent {\bf Author Contribution:} H.K. and Y.C. fabricated samples with the
help of Y.Z. and H.Z., and performed STM measurements. H.K., Y.C., and S.N.-P.
analyzed the data with the help of L.K. and E.B. E. L.-H., C.L. and A.T. provided 
the theoretical analysis supervised by J.A. S.N.-P. supervised the project. 
H.K., Y.C., E. L.-H., C.L., A.T., J.A., and S.N.-P. wrote the manuscript with 
input from other authors.

\noindent {\bf Data availability:} The data that support the findings of this
study are available from the corresponding authors on reasonable request.

\clearpage

\clearpage

\clearpage

\clearpage

\renewcommand{\figurename}{\textbf{Extended Data Fig.}}
\renewcommand{\theHfigure}{Extended.\thefigure}
\setcounter{figure}{0}

\begin{figure}[p]
\begin{center}
    \includegraphics[width=15cm]{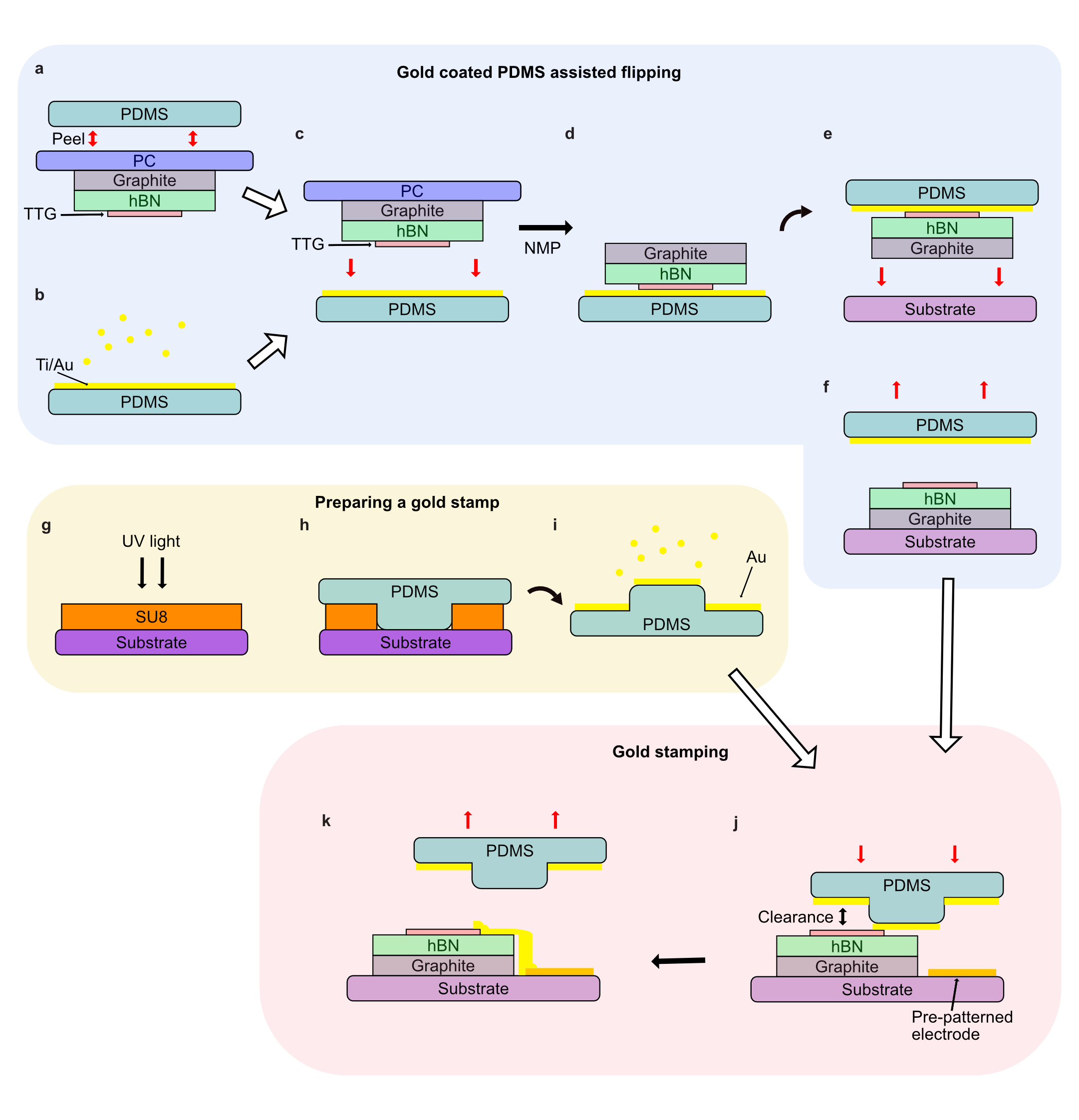}
\end{center}
\caption{{\bf Sample fabrication.}
{\bf a-f}, gold coated PDMS assisted flipping. A stack with PC film is peeled off from a PDMS block ({\bf a}). A separate PDMS block is Ti/Au coated ({\bf b}). Then the stack is put down to the gold coated PDMS ({\bf c}). PC film is dissoved by NMP ({\bf d}), before the stack is dropped down to a substrate ({\bf e,f}). {\bf g-i}, a gold stamp is prepared. A mold is defined by photolithography ({\bf g}). PDMS is poured on the mold ({\bf h}), and peeled off. Au is deposited on the stamp ({\bf i}).
{\bf j,k}, the gold stamp is pressed down onto a desired area of the sample ({\bf j}), leaving a gold strip that connects the sample and a pre-patterned electrode on the substrate ({\bf k}). 
}
\label{exfig: Method}
\end{figure}

\begin{figure}[p]
\begin{center}
    \includegraphics[width=15cm]{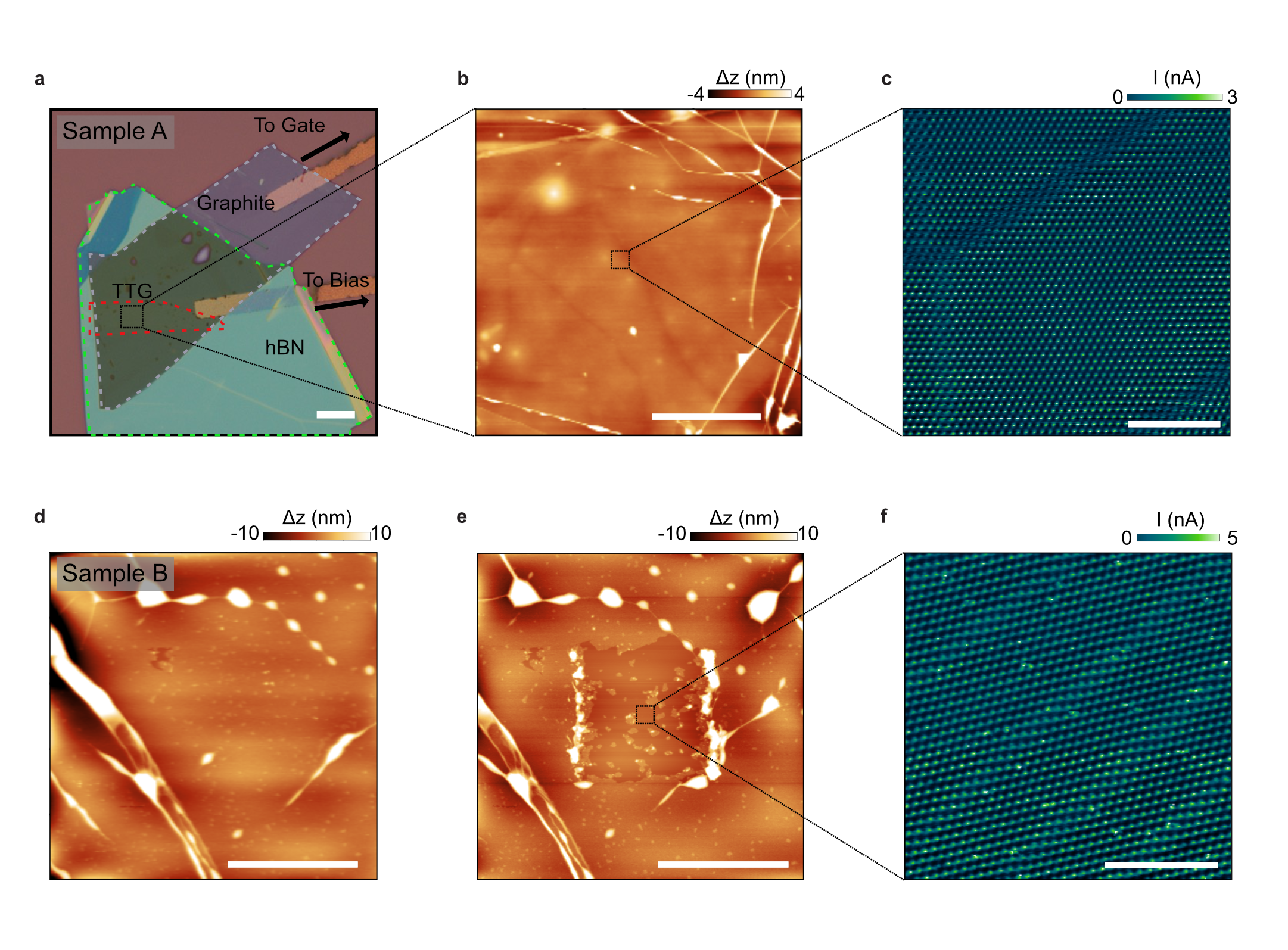}
\end{center}
\caption{{\bf AFM images of samples A and B.}
{\bf a}, Optical microscope image of Sample A. {\bf b}, AC tapping mode AFM image after the contact mode cleaning the $2 \times 2 \mu m^{2}$ area around the center of the image. No sign of residue is found. {\bf c}, cAFM image showing moir\' e pattern of MATTG. {\bf d,e}, AC tapping mode AFM images of Sample B before ({\bf d}) and after ({\bf e}) cleaning. The residue boundaries after the cleaning indicate significant amount of residues on the surface. {\bf f}, cAFM image.
Scale bars : $ 10 \mu m $ ({\bf a}), $ 2 \mu m $ ({\bf b,d,e}), $ 100 nm $ ({\bf c,f}). 
}
\label{exfig: AFM}
\end{figure}

\begin{figure}[p]
\begin{center}
    \includegraphics[width=15cm]{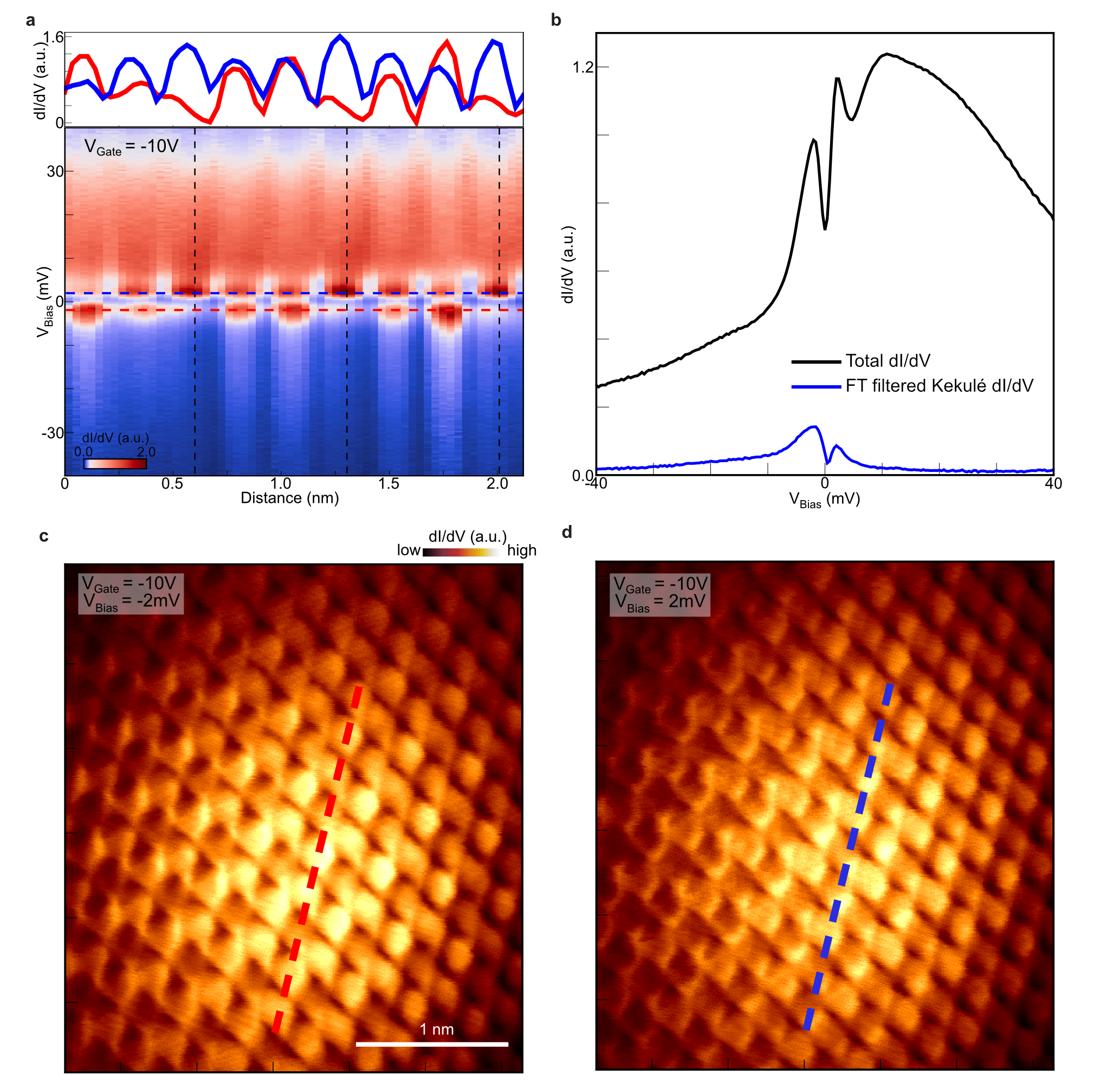}
\end{center}
\caption{{\bf $\mathbf{V_{\rm Bias}}$ dependent mapping of the lattice tripling order on MATTG.}
{\bf a}, Conductance at $V_{\rm Gate} = -10$V taken along spatial points and for range of $V_{\rm Bias}$. While at large positive $V_{\rm Bias}$, $dI/dV$ shows periodic modulation that corresponds to graphene lattices, at low $V_{\rm Bias}$, additional periodic pattern that triples the graphene lattice periodicity is apparent. (black dashed lines) Upper inset shows two linecut taken at above ($V_{\rm Bias} = 2$mV) and below ($V_{\rm Bias} = -2$mV) $E_F$ stressing lattice tripling. {\bf b}, $V_{\rm Bias}$ spectroscopy extracted from \prettyref{exfig: Kekule_bias}a that compares total $dI/dV$ obtained by summing up along spatial coordinates and FT filtered Kekul\'e $dI/dV$ which is a result of FT filtering on \prettyref{exfig: Kekule_bias}a along spatial direction to extract Kekul\'e signal. {\bf c,d}, $dI/dV$ map measured at fixed $V_{\rm Gate} = -10$V at negative $V_{\rm Bias} = -2$mV ({\bf c}) and positive $V_{\rm Bias} = 2$mV ({\bf d}). Red and blue dashed line shows the spatial positions where \prettyref{exfig: Kekule_bias}a is measured. Measurements are taken at $T = 400$mK.}
\label{exfig: Kekule_bias}
\end{figure}

\clearpage

\begin{figure}[p]
\begin{center}
    \includegraphics[width=15cm]{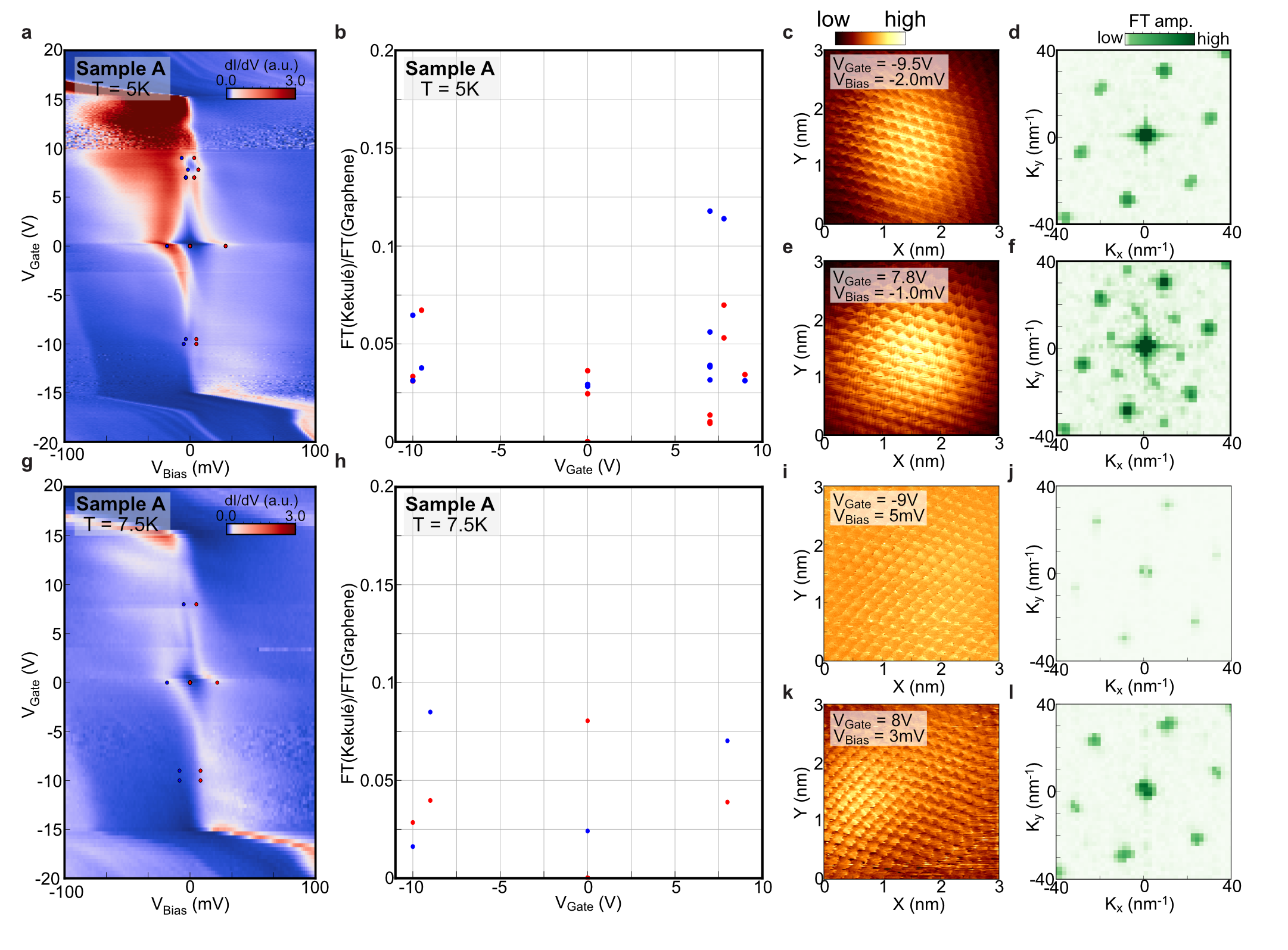}
\end{center}
\caption{{\bf $\mathbf{V_{\rm Gate}}$ dependent evolution of the lattice tripling order at higher temperatures.}
{\bf a},  $V_{\rm Gate}$ dependent $dI/dV$ spectroscopy measured at $T = 5$K where $\nu = 2$ correlated gap survives but gaps around $\nu = -2$ is greatly suppressed. Red (positive $V_{\rm Bias}$) and Blue (negative $V_{\rm Bias}$) dots marks the position where we measured 2D $dI/dV$ maps. {\bf b}, Intensity of the peak at Kekul\'e reciprocal lattice vector normalized by the intensity of the peak at graphene reciprocal lattice vector as a function of $\mathbf{V_{Gate}}$. {\bf c,e} Real space $dI/dV$ map at $V_{\rm Gate} = -9.5$V ({\bf c}) and $V_{\rm Gate} = 7.8$V ({\bf e}). {\bf d,f} Fourier transformation of \prettyref{exfig: Kekule_T}c,e.
{\bf g}, $V_{\rm Gate}$ dependent $dI/dV$ spectroscopy measured at $T = 7.5$K where $\nu = 2$ correlated gap survives but gaps around $\nu = -2$ are greatly suppressed. Red (positive $V_{\rm Bias}$) and Blue (negative $V_{\rm Bias}$) dots mark the position where we measured 2D $dI/dV$ maps. {\bf h}, Intensity of the peak at Kekul\'e reciprocal lattice vector normalized by the intensity of the peak at graphene reciprocal lattice vector as a function of $\mathbf{V_{Gate}}$. {\bf i,j} Real space $dI/dV$ map at $V_{\rm Gate} = -9.5$V ({\bf i}) and $V_{\rm Gate} = 7.8$V ({\bf j}). {\bf k,l} Fourier transformation of \prettyref{exfig: Kekule_T}k,l. }
\label{exfig: Kekule_T}
\end{figure}

\clearpage

\begin{figure}[p]
\begin{center}
    \includegraphics[width=15cm]{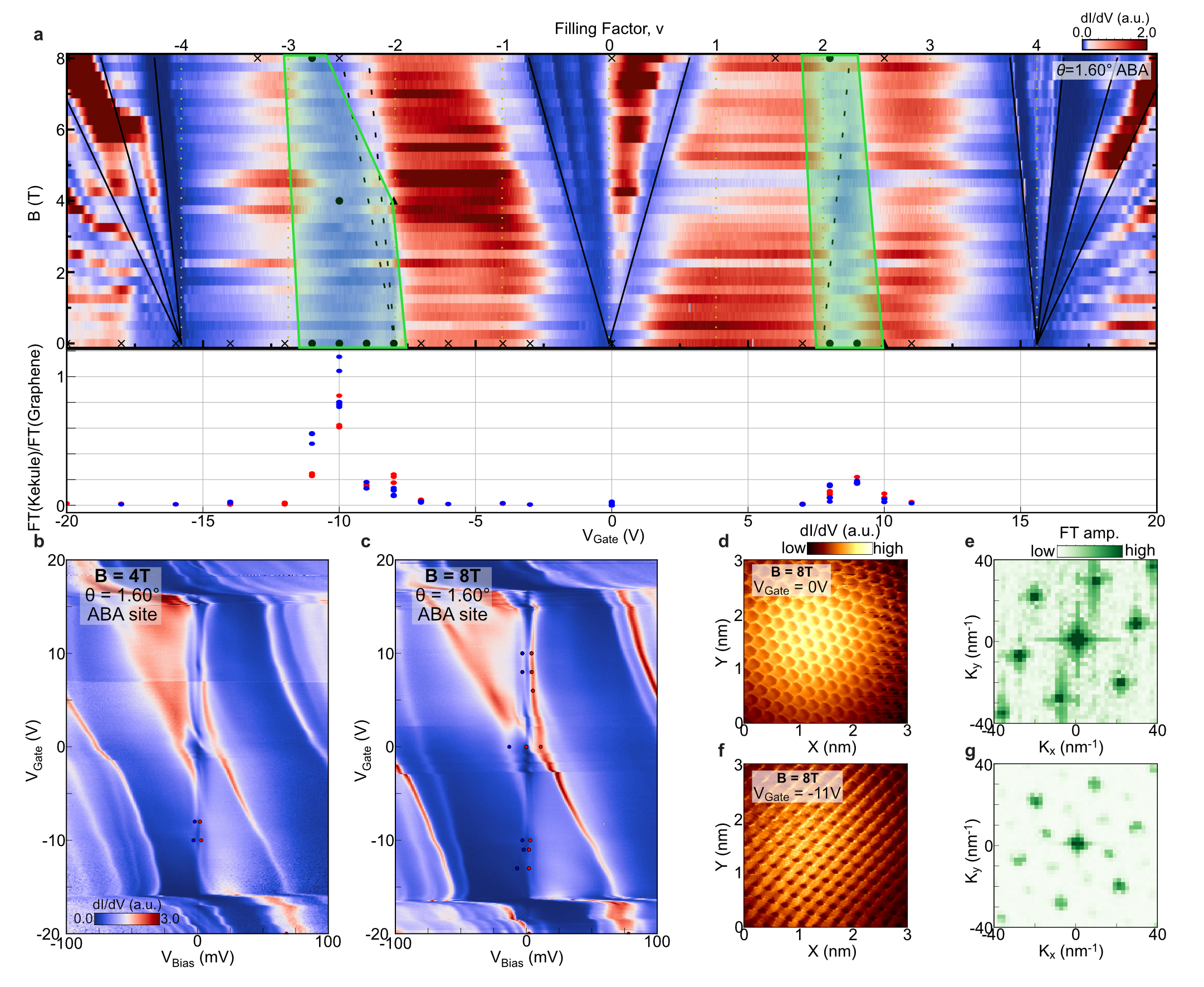}
\end{center}
\caption{{\bf Out-of-plane magnetic field dependence of lattice tripling order.}
{\bf a}, LDOS Landau fan diagram measured on an moir\'e ABA site. The lower panel shows intensity of the lattice tripling signal as in \prettyref{fig: fig2}. Black circles marked on Landau fan indicate $V_{\rm Gate}$ and $\mathrm{B}$, where we observe Kekul\'e peaks in FT. The black cross indicate values of $V_{\rm Gate}$ and $\mathrm{B}$ where we measured $dI/dV$ map but could not observe Kekul\'e peaks in FT. The green polygon is an eye guide covering black circles and roughly denotes where we observed lattice tripling.
{\bf b, c},  $V_{\rm Gate}$ dependent $dI/dV$ spectroscopy measured at $B = 4$T ({\bf b}) and $B = 8$T ({\bf c}).
{\bf d, e}, Real space $dI/dV$ map ({\bf d}) and corresponding Fourier transformation ({\bf e}) showing the absence of Kekul\'e FT peaks around CNP at $B = 8$T.
{\bf f, g}, Real space $dI/dV$ map ({\bf f}) and corresponding Fourier transformation ({\bf g}) taken at $B = 8$T that shows Kekul\'e order at $V_{\rm Gate} = -11$V.
}
\label{exfig: Kekule_B}
\end{figure}

\clearpage

\begin{figure}[p]
\begin{center}
    \includegraphics[width=15cm]{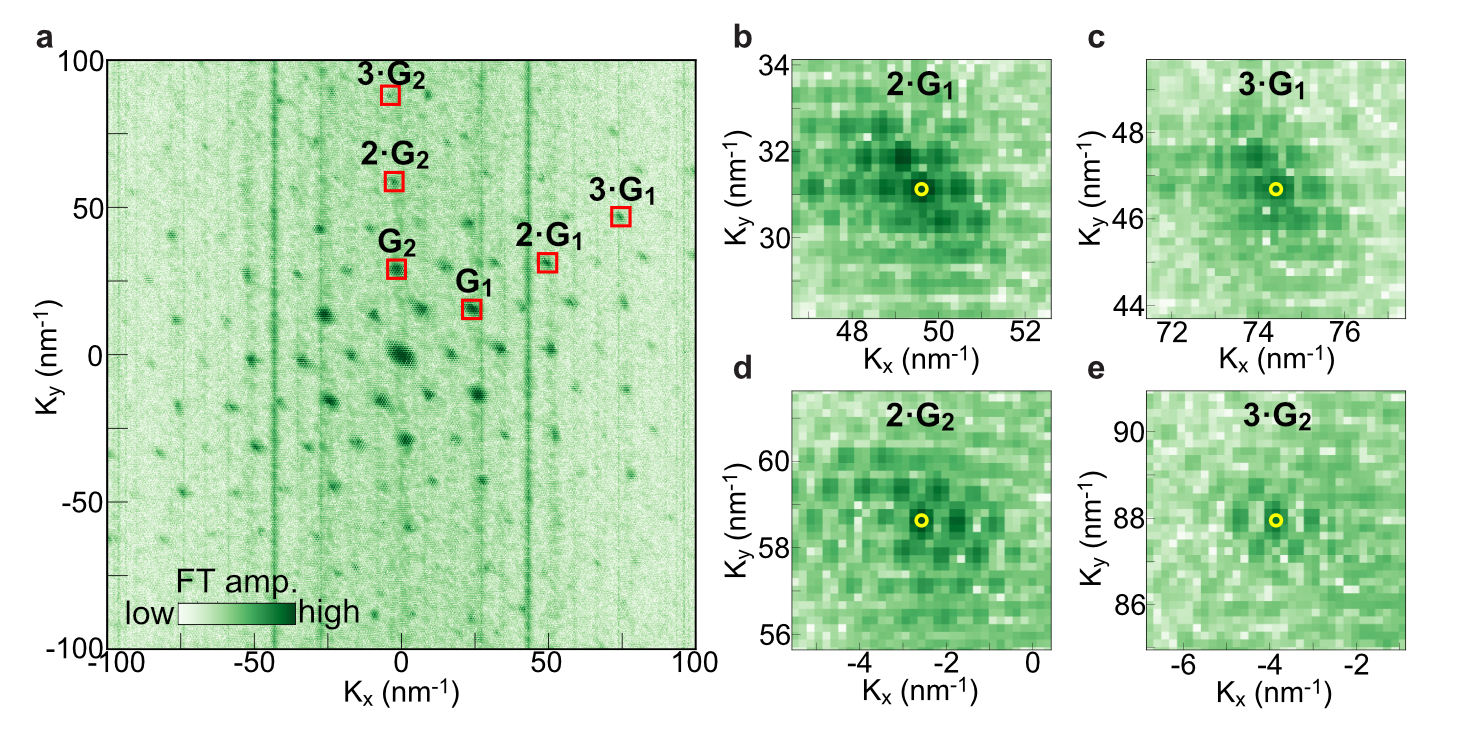}
\end{center}
\caption{{\bf High-order graphene reciprocal lattice vector peaks from FT map at $\mathbf{\nu = -2.3}$.}
{\bf a}, Fourier transformation of the real space $dI/dV$ map at $V_{\rm Gate} = -9$V and $V_{\rm Bias} = -2$mV showing larger momentum range compared to \prettyref{fig: fig4}a.
{\bf b-e}, Zoom-in of the \prettyref{exfig: Higher_order_peaks}a around $2\bG_{1}$ ({\bf b}), $3\bG_{1}$ ({\bf c}), $2\bG_{2}$ ({\bf d}), $3\bG_{2}$ ({\bf e}) that is marked as a red rectangle in \prettyref{exfig: Higher_order_peaks}a. Position of $N\bG_{i}$ that is determined from the positions of $\bG_{1}$ and $\bG_{2}$ from \prettyref{fig: fig4}b,c is plotted as yellow circles which matches well with the FT peaks.
}
\label{exfig: Higher_order_peaks}
\end{figure}

\clearpage

\begin{figure}[p]
\begin{center}
    \includegraphics[width=15cm]{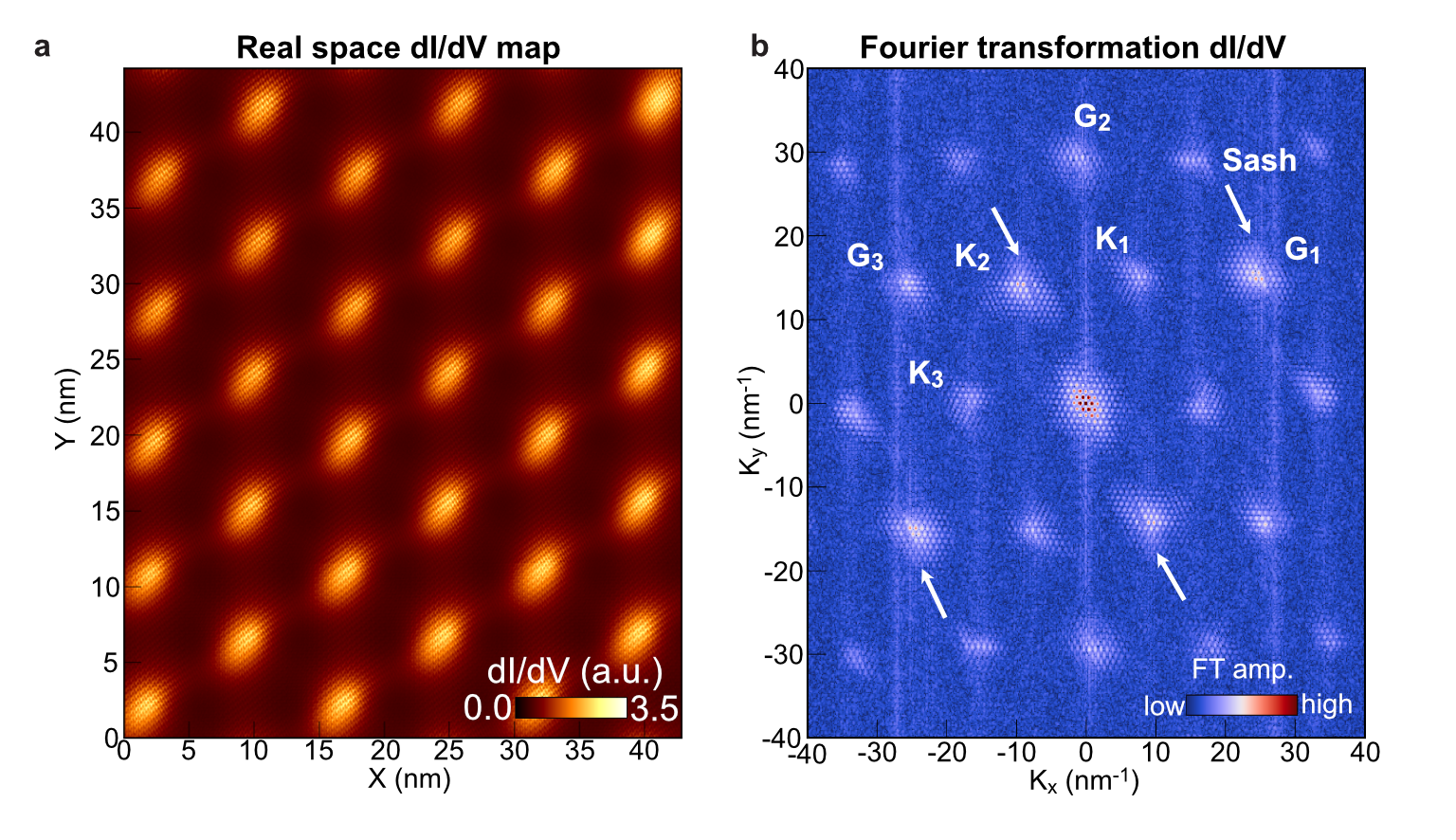}
\end{center}
\caption{{\bf 42 nm by 42 nm size $dI/dV$ map with Fourier transformation at $\mathbf{\mu = -2.3}$ showing sash features.}
{\bf a,b}, Real space $dI/dV$ map ({\bf a}) and corresponding Fourier transformation ({\bf b}) taken at $V_{\rm Bias} = -2$mV. White arrows point to the 'sash' features.}
\label{exfig: sash_features}
\end{figure}

\clearpage

\begin{figure}[p]
\begin{center}
    \includegraphics[width=15cm]{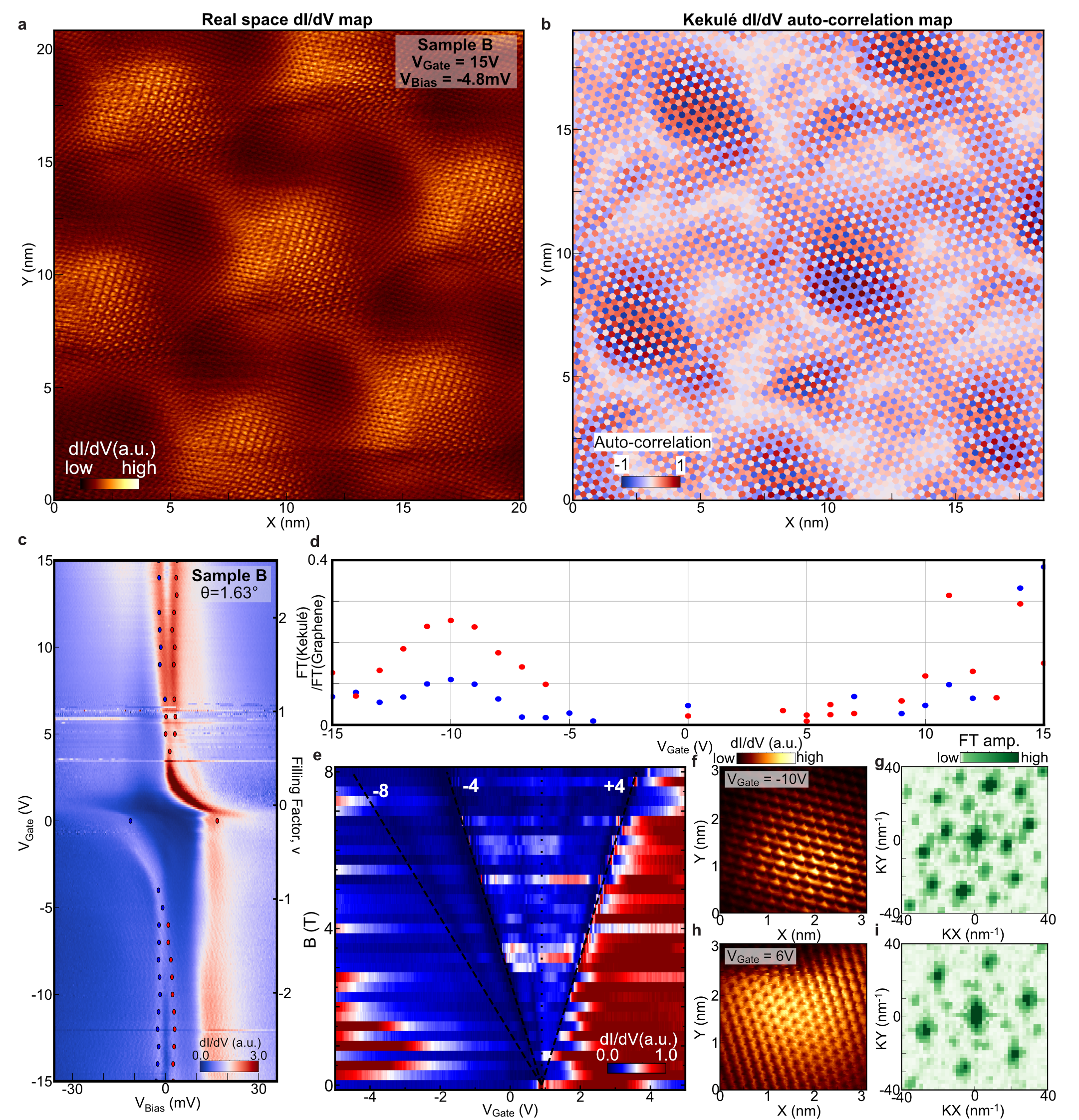}
\end{center}
\caption{{\bf Lattice tripling order observed in sample B with higher heterostrain.}
{\bf a}, Real space $dI/dV$ map taken at $V_{\rm Gate} = 15$V and $V_{\rm Bias} = -4.8$mV that includes seven moir\'e AAA sites. 
{\bf b}, Kekul\'e auto-correlation map created from real space $dI/dV$ map in \prettyref{exfig: High_strain_sample}a. Neighboring AAA sites are mapped with different colors, exhibiting the change in Kekul\'e patterns.
{\bf c}, $V_{\rm Gate}$ dependent $dI/dV$ spectroscopy measured on sample B. {\bf d}, $V_{\rm Gate}$ dependence of the Kekul\'e peak intensity in FT images normalized by the graphene lattice peak. Red (Blue) dot corresponds to positive (negative) $V_{\rm Bias}$, and is marked in \prettyref{exfig: High_strain_sample}c.
{\bf e}, LDOS Landau fan diagram measured on sample B. Landau level degeneracies at each insulating dip is written in white numbers. 
{\bf f, g}, Real space $dI/dV$ map ({\bf f}) and Fourier transformation ({\bf g}) taken at $V_{\rm Gate} = -10$V showing lattice tripling.
{\bf h, i}, Real space $dI/dV$ map ({\bf h}) and Fourier transformation ({\bf i}) taken at $V_{\rm Gate} = 19$V. Measurements are taken at $T = 2$K.
}
\label{exfig: High_strain_sample}
\end{figure}

\clearpage

\begin{figure}[p]
\begin{center}
    \includegraphics[width=15cm]{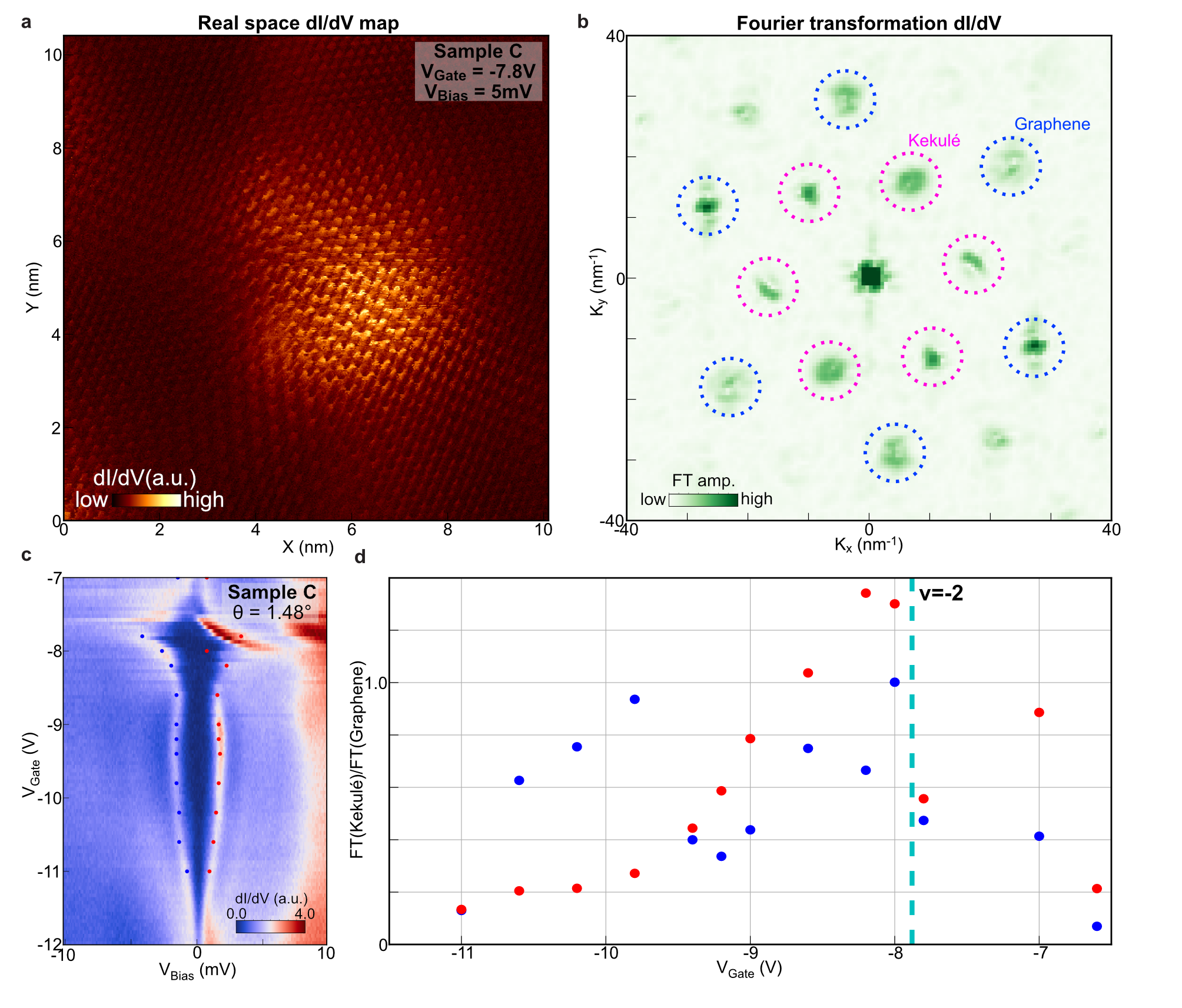}
\end{center}
\caption{{\bf Lattice tripling order observed in sample C with lower heterostrain.}
{\bf a}, Real space $dI/dV$ map on one moir\'e AAA site measured at $V_{\rm Gate} = -7.8$V and $V_{\rm Bias} = 5$mV which is at $\nu = -2$.
{\bf b}, Fourier transformation of \prettyref{exfig: Sc_sample}a exhibiting prominent Kekul\'e FT peaks.
{\bf c}, $V_{\rm Gate}$ dependent $dI/dV$ spectroscopy focusing on the correlated gaps at $\nu = -2\sim-3$.
{\bf d}, $V_{\rm Gate}$ dependence of the Kekul\'e peak intensity in FT images normalized by the graphene lattice peak. Measurements are taken at $T = 400$mK.
}
\label{exfig: Sc_sample}
\end{figure}

\clearpage

\begin{figure}[p]
\begin{center}
    \includegraphics[width=15cm]{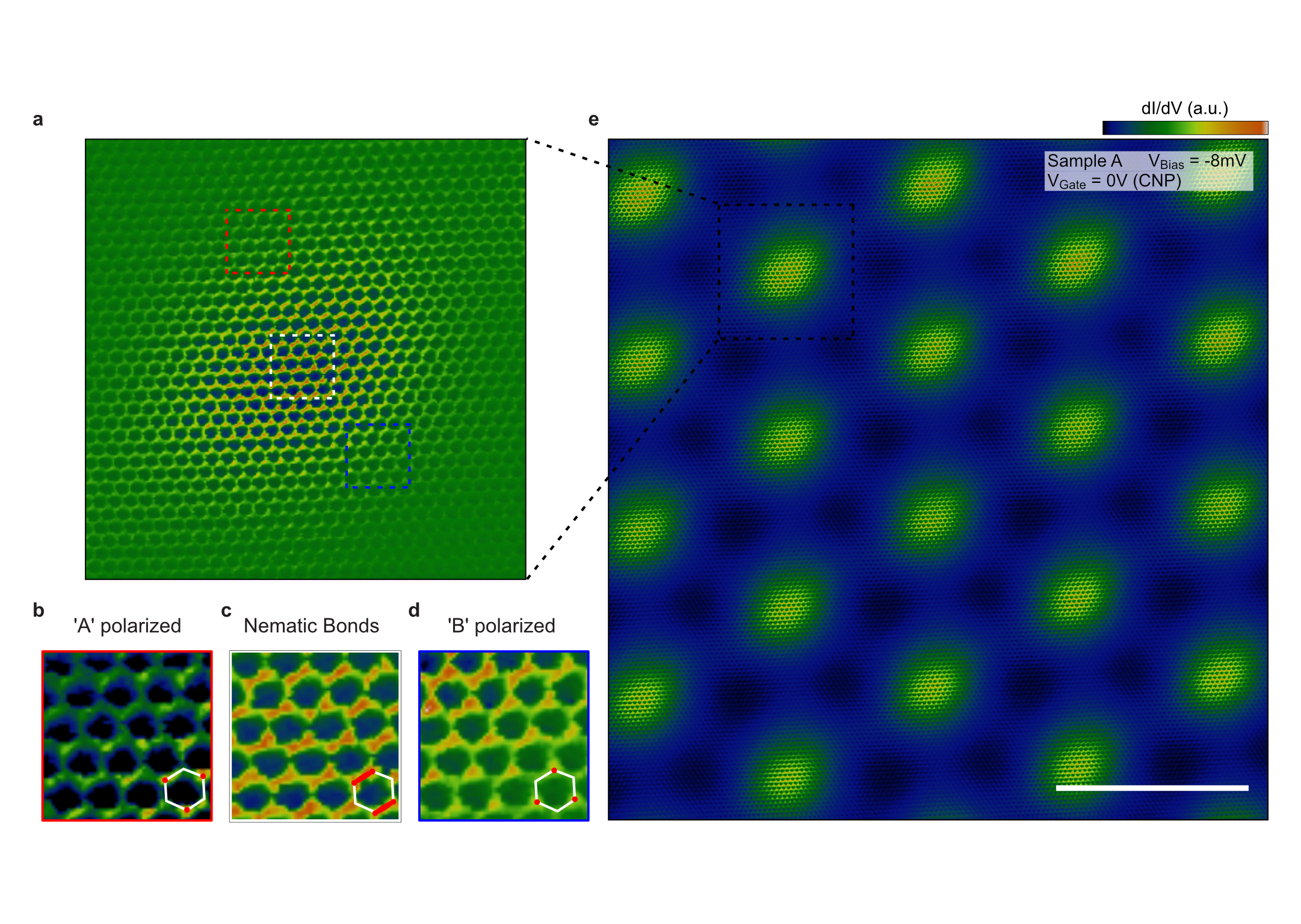}
\end{center}
\caption{{\bf Nematic semimetal phase at charge neutrality.}
{\bf a}, atomic resolution dI/dV map (background filtered) at an AAA site. 
{\bf b-d}, signatures of nematic seminetallic (NSM) ground state. The intensity of one bond is stronger than the other two bonds near the center of the AAA site ({\bf c}). The pattern slowly evolves into 'A' sublattice polarized state around the red boxed region ({\bf b}), while 'B' sublattice polarized state is dominant around the blue boxed region ({\bf d}), which agrees the prediction in Ref.~\citenum{hongDetectingSymmetryBreaking2022}. 
{\bf e}, (unprocessed) dI/dV map at $V_{\rm Gate} = 0V$ (CNP) on the hole side ($V_{\rm Bias} = -8mV$) where ({\bf a-d}) are taken. Other AAA sites in this map also show similar behavior as ({\bf a-d}). This map and the map in Fig.~\ref{fig: fig3} are taken at the same area, only having different gate and bias voltages. Scale bar : 10 nm.
}
\label{exfig: nsm}
\end{figure}

\clearpage

\clearpage
\beginsupplement
\renewcommand{\theHfigure}{Supplement.\thefigure}
\noindent \textbf{Supplementary Information: Imaging inter-valley coherent order in magic-angle twisted trilayer graphene\\ 
}

\vspace{5pt}

{\small Hyunjin Kim, Youngjoon Choi, \'Etienne Lantagne-Hurtubise, Cyprian Lewandowski, Alex Thomson, Lingyuan Kong, Haoxin Zhou, Eli Baum,
Yiran Zhang, Ludwig Holleis, Kenji Watanabe, Takashi Taniguchi, Andrea F. Young, Jason Alicea, Stevan Nadj-Perge}

\vspace{5pt}

This Appendix provides details of the theoretical modeling supporting this work and is organized as follows. We first establish conventions and review the continuum model for twisted graphene superlattices in Sec.~\ref{app:BM}. We then review the treatment of heterostrain and its profound effects on the low-energy (flat) bands of MATTG in Sec.~\ref{app:heterostrain}. In Sec.~\ref{app:Hartree}, we investigate Hartree corrections to the MATTG flat bands in the presence of strain, focusing on filling factors between $-3$ and $-2$. In Sec.~\ref{app:strained_bands}, we explore various types of inter-valley coherent instabilities in MATTG. In particular, we extract trends related to the preferred IKS wavevector as a function of strain parameters, interaction strength, and the chiral ratio in the BM model.

In Sec.~\ref{app:modeling_IVC} we introduce more formally different inter-valley coherent orders, and in Sec.~\ref{app:LDOS_method} we present an efficient numerical method to compute real-space lattice tripling signatures in the local density of states (LDOS). We contrast the implications for moir\'e-periodic IVC orders to those of incommensurate Kekul\'e spiral (IKS) states in Sec.~\ref{app:LDOS_results}. Finally, in Sec.~\ref{app:comensurate}, we speculate on a scenario whereby the incommensurate Kekul\'e spiral might undergo a lock-in mechanism to a nearby commensurate wavevector.

\section{Conventions, geometry and the BM model}
\label{app:BM}

We first establish our conventions. We denote the carbon-carbon length $a_0 = 0.142$ nm. We take the Bravais lattice vectors  of monolayer graphene as
\begin{align}
\bA_1 =& \sqrt{3} a_0 \left( \frac{1}{2}, \frac{\sqrt{3}}{2} \right) ~ , ~
\bA_2 = \sqrt{3} a_0 \left( \frac{1}{2}, -\frac{\sqrt{3}}{2} \right),
\end{align}
and its reciprocal lattice vectors
\begin{align}
\bG_1 =& \frac{4 \pi}{3 a_0} \left( \frac{\sqrt{3}}{2}, \frac{1}{2} \right) ~ , ~
\bG_2 = \frac{4 \pi}{3 a_0} \left( \frac{\sqrt{3}}{2}, -\frac{1}{2} \right) .
\end{align}
The graphene Dirac points are located at $\bK_{\pm} = \mp \frac{4 \pi}{3 \sqrt{3} a_0} (1, 0)$. (In the main text, we refer alternatively to the two Dirac points as $\bK$ and $\bK'$.)

We consider alternating-twist trilayer graphene (TTG)\cite{khalafMagicAngleHierarchy2019, carrUltraheavyUltrarelativisticDirac2020} with a twist angle $-\theta/2$ (i.e., in the clockwise direction) applied to the top and bottom layers, and $\theta/2$ (i.e., in the counterclockwise direction) applied to the middle layer. The corresponding moir\'e pattern is characterized by Bravais lattice vectors
\begin{align}
\ba_1 =& \frac{4 \pi}{3 k_\theta} \left( \frac{\sqrt{3}}{2}, \frac{1}{2} \right) ~ , ~
\ba_2 =  \frac{4 \pi}{3 k_\theta} \left( 0, 1 \right),
\end{align}
and reciprocal lattice vectors
\begin{align}
\bg_1 =& \sqrt{3} k_\theta \left( 1, 0 \right) ~ , ~
\bg_2 = \sqrt{3} k_\theta \left( -\frac{1}{2}, \frac{\sqrt{3}}{2} \right),
\end{align}
with $k_\theta = 2 | \bK_\pm | \sin \frac{\theta}{2}$ the momentum scale between the Dirac points originating from (say) the top and middle layers. In the MATTG sample investigated in the main text, the twist angle $\theta = 1.602^\circ$ translates to a moir\'e lengthscale $l_M = 4 \pi / 3 k_{\theta} \approx 8.80$ nm.

To describe the low-energy physics of the system our starting point is the continuum Bistritzer-MacDonald (BM) model~\cite{Lopes2007, bistritzerMoireBandsTwisted2011}, which consists of intra-layer kinetic energy contributions as well as inter-layer tunneling modulated by the moir\'e potential. We treat the spin degree of freedom as a spectator throughout. In the low-energy (linearized) approximation for the original Dirac cones of each graphene layer, the intra-layer contribution can be written as
\begin{align}
h^l_\tau(\bk, \theta) & = \hbar v_F [R^T(\theta_l) (\gamma_\tau + \bk - \bK_{\tau,l}) ] \cdot (\tau \sigma_x, \sigma_y)
\label{eq:intralayer}
\end{align}
where $l=1,2,3$ denotes the layer and $\tau = \pm$ the valley, $\sigma_x, \sigma_y$ are Pauli matrices acting on the sublattice (A/B) degree of freedom, $R(\theta)$ is the usual counter-clockwise rotation matrix and $v_F \sim 8.7 \times 10^5$ m/s is an effective Fermi velocity, similar to the value for monolayer graphene (but larger than the value obtained from fitting to ab initio calculations~\cite{koshinoMaximallyLocalizedWannier2018} in TBG).The Dirac point in layer $l$ is rotated as $\bK_{\tau,l} = R(\theta_l) \bK_\tau$. The momentum $\gamma_\tau + \bk$ in Eq.~\ref{eq:intralayer} denotes the microscopic momentum of each graphene layer, measured with respect to the original ${\bm \Gamma}$ point, while $\bk$ is defined from the mBZ center in each valley, $\gamma_{\tau} = -\tau \frac{4 \pi}{3 \sqrt{3} a_0} \left( \sqrt{3} \sin( \frac{\theta}{2} ) + \cos( \frac{\theta}{2} ), 0 \right)$, see Fig.~\ref{fig:IKS}a. 

The inter-layer tunneling between adjacent graphene layers is modulated by the moir\'e potential, and reads
\begin{align}
    T(\br) = T_1 + T_2 e^{i (\bg_1 + \bg_2) \cdot \br} + T_3 e^{i \bg_2 \cdot \br},
\end{align}
where the three matrices
\begin{align}
T_1 = \begin{pmatrix}
    w_0 & w_1 \\
    w_1 & w_0 
\end{pmatrix}, ~
T_2 = \begin{pmatrix}
    w_0 & w_1 e^{-i  \phi}\\
    w_1 e^{i \phi} & w_0
\end{pmatrix}, ~
T_3 = \begin{pmatrix}
    w_0 & w_1 e^{i \phi} \\
    w_1 e^{-i \phi} & w_0
\end{pmatrix},
\label{eq:Tj_definition}
\end{align}
act in the sublattice space, and the phase factor $\phi = 2 \pi/ 3$. 
In valley $\tau=-$, the inter-layer tunneling terms are obtained by time-reversal symmetry which (in the spinless version of the problem) flips all momenta and takes $\phi \rightarrow -\phi$.
For the tunneling parameters we take $\omega_0 = 55$ meV and $\omega_1 = 105$ meV, which produces a gap of $\sim 80 - 90$ meV between the flat bands and the higher-energy bands, depending on the strength of interactions (see Sec.~\ref{app:Hartree} and also Ref.~\onlinecite{kimEvidenceUnconventionalSuperconductivity2022}), similar to that observed in the experiment. The corresponding ratio $\eta = w_0/w_1 \approx 0.52$ is closer to the chiral limit~\cite{tarnopolskyOriginMagicAngles2019} $\eta = 0$ than reported in recent \emph{ab initio} studies~\cite{zhuTwistedTrilayerGraphene2020, carrUltraheavyUltrarelativisticDirac2020} of MATTG. 

As shown in Ref.~\citenum{khalafMagicAngleHierarchy2019}, twisted trilayer graphene possesses a mirror symmetry that interchanges the top and bottom layers. Using the eigenbasis of the mirror symmetry operator
\begin{equation}
 \cM_z = \begin{pmatrix}
     0 & 0 & \mathds{1}_{2 \times 2} \\
     0 & \mathds{1}_{2 \times 2} & 0 \\
     \mathds{1}_{2 \times 2} & 0 & 0
 \end{pmatrix}    
\end{equation}
acting in the layer and sublattice space spanned by $(A_1, B_1, A_2, B_2, A_3, B_3)$, the BM Hamiltonian can be block-diagonalized to an odd-parity subspace (comprising the anti-symmetric combination of layers $1$ and $3$) and an even-parity subspace (comprising layer $2$ and the symmetric combination of layers $1$ and $3$). The odd sector contributes a (rotated) Dirac cone inherited from the monolayer graphene dispersion, whereas the even subspace is analogous to twisted bilayer graphene but with an ``effective" tunneling strength enhanced by a factor of $\sqrt{2}$. We can therefore estimate the magic angle of MATTG from the parameters defined above, using the condition $\alpha = w_1 / (v_F k_{\theta}) \approx 0.6$ from MATBG~\cite{bistritzerMoireBandsTwisted2011} but multiplying the tunneling parameter $w_1$ by $\sqrt{2}$ to account for the symmetry transformation relating MATTG to MATBG, yielding $\theta_M \sim 1.52^\circ$.

In the presence of a perpendicular displacement field $D$, an interlayer potential $u = - d D/{\epsilon_\perp}$ is generated between neighboring layers, where $d \approx 0.33$ nm is the interlayer distance and $\epsilon_{\perp}$ the dielectric constant of the device in the perpendicular direction. Such a term breaks the mirror symmetry decomposition and hybridizes the flat bands with the Dirac cone.

\section{Heterostrain}
\label{app:heterostrain}

The TTG samples studied in this work are subject to heterostrain that acts in an opposite way between adjacent rotated layers (a type of strain arising from lattice relaxation within the moir\'e unit cell and instrinsic to the sample, in contrast to strain inherited e.g. from a substrate). The combination of heterostrain and rigid twist angle has profound consequences for the low-energy flat bands~\cite{biDesigningFlatBands2019} as well as for the phase diagram in the presence of electronic interactions~\cite{parkerStrainInducedQuantumPhase2021}. This can be understood by noting that the strain energy scale $\epsilon \hbar v_F / a_0$ amounts to a few meV (and is thus comparable to the bandwidth of the moir\'e bands) for experimentally-relevant $\epsilon \sim 0.1 - 0.2\%$. Therefore, from the perspective of the low-energy bands the $C_{3z}$ symmetry is strongly broken. Heterotrain however preserves the $C_{2z}$ symmetry of pristine TTG. 

We characterize the heterostrain following Ref.~\citenum{biDesigningFlatBands2019}. We define the deformation tensor $\cE_l$ in layer $l$, which includes strain in addition to the uniform rotation by the twist angle $\theta_l = (-1)^{l} \theta/2$, as
\begin{align}
\cE_l &= R^T(\varphi_l) 
\begin{pmatrix}
-\epsilon_l & 0 \\
0  & \nu \epsilon_l
\end{pmatrix}
 R(\varphi_l) + R(\theta_l) - \mathds{1} \\
 &\approx 
\begin{pmatrix}
\epsilon^l_{xx} & \epsilon^l_{xy} -\theta_l \\
\epsilon^l_{xy} +\theta_l & \epsilon^{l}_{yy}
\end{pmatrix}.
\end{align}
Here $\epsilon_l$ is the magnitude of strain in layer $l$ and $\varphi_l$ its direction (i.e., the angle of the compressed axis with respect to the $x$ axis). The parameter $\nu \approx 0.16$ is the Poisson ratio for monolayer graphene. The strain tensor components are denoted by $\epsilon^l_{ij}$. 
We further assume that strain acts in an identical way on the top and bottom layers (but with an opposite sign on the middle layer), i.e., $\epsilon_{1} = \epsilon_3 = -\epsilon_2 \equiv \epsilon$ and $\varphi_1 = \varphi_2 = \varphi_3 \equiv \varphi$. Deviations from this condition are expected to lead to much weaker effects on the band structure, because the top and bottom layers are not twisted relative to one another.  (In contrast, the effect of heterostrain between layers $1$ or $3$ and layer $2$ is magnified due to their small relative twist angle.) This strain configuration preserves the mirror symmetry $\cM_z$.

In real and momentum space, vectors on the monolayer graphene scale transform (assuming zero displacement between the layers) as
\begin{equation}
\br_l \rightarrow \cM_l \br_l ~ , ~ \bk_l \rightarrow (\cM_l^T)^{-1} \bk_l ,
\end{equation}
with $\cM_l = 1 + \cE_l$. These transformations preserve the inner product between Bravais and reciprocal lattice vectors.
The mBZ can be constructed as follows. The moiré reciprocal lattice vectors are obtained from the subtraction of the deformed reciprocal lattice vectors of monolayer graphene in the different layers as
\begin{align}
\bg_1 =& \left( (\cM_1^T)^{-1} - (\cM_2^T)^{-1} \right) \left( \bG_1- \bG_2 \right) , \\
\bg_2 =&  \left( (\cM_1^T)^{-1} - (\cM_2^T)^{-1} \right) \left( - \bG_1 \right).
\end{align}
The resulting mBZ geometry is shown in Fig.~\ref{fig:IKS}a. Note that the center of the strained moir\'e BZ, $\gamma_{\tau} = \frac{1}{2} \left( \bK_{\tau, 1} + \bK_{\tau, 2} \right) - \frac{ \bg_1}{2}$, is also displaced slightly
from its original location. We also define a third moir\'e reciprocal lattice vector, $\bg_3 = - (\bg_1 + \bg_2)$, for future convenience. The strained moir\'e Bravais lattice vectors, $\ba_i$ with $i=1,2$, can be obtained from the $\bg_j$ through the defining relation $\ba_i \cdot \bg_j = 2 \pi \delta_{ij}$. The parameters $\epsilon$ and $\varphi$ characterizing the experimental geometry are obtained by reproducing the strained Bravais lattice vectors  extracted from large-area topography measurements (see also Methods section). This analysis leads to $\epsilon \approx -0.12 \%$ (i.e., the middle layer is compressed while the top/bottom layers expands along the $\varphi$ direction) with $\varphi \approx 87^\circ$.

In the presence of heterostrain, the BM continuum model for TTG is affected in two ways. The first effect is geometric: the strain changes the shape of the mBZ according to the transformations outlined above. The second effect, inherited from the coupling of the underlying Dirac electrons of monolayer graphene to the strain field, is the emergence of a ``pseudo" vector potential

\begin{equation}
\bA_l = \frac{\beta}{2} ( \epsilon^l_{xx} - \epsilon^l_{yy}, -2 \epsilon^l_{xy} ),
\end{equation}
which couples to the microscopic momentum $\bk$ through minimal substituation, $\bk \rightarrow \bk + \tau \bA$. Here $\beta \approx 3.14$ relates the change in hopping energy to the change in distance between orbitals in monolayer graphene. The different sign of the mininal substitution between the two valleys is mandated by time-reversal symmetry.

In terms of the strain angle $\varphi_l$ and magnitude $\epsilon_l$, the strain tensor components
\begin{align}
\epsilon^l_{xx} &= \epsilon_l (\nu \sin^2\varphi_l - \cos^2\varphi_l) , \\
\epsilon^l_{yy} &= \epsilon_l (\nu \cos^2\varphi_l - \sin^2\varphi_l ),\\
\epsilon^l_{xy} &= (1 + \nu) \epsilon_l \cos \varphi_l \sin \varphi_l,
\end{align}
are manifestly invariant under $\varphi_l \rightarrow \varphi_l + \pi$. The pseudo-vector potential
\begin{align}
\bA_l %
&= - \frac{\beta \epsilon_l}{2} (\nu + 1) (\cos 2 \varphi_l, \sin 2 \varphi_l)
\end{align}
has an additional symmetry: it is invariant under $\varphi_l \rightarrow \varphi_l + \pi/2$ followed by $\epsilon_l \rightarrow - \epsilon_l$. This last transformation however changes the geometry of the mBZ, as the Poisson ratio $\nu \neq 1$. This observation allows us to determine both the magnitude and sign of $\epsilon$ in the experiment (see also the Methods section).

In the presence of heterostrain, only $\cC_{2z}$, $T$ and the mirror $\cM_z$ (if the displacement field $D=0$) remain symmetries of the problem. When diagonalizing the BM model, we enforce that its eigenfunctions  respect $\cC_{2z} \cal T$ symmetry, acting as $\left( \cC_{2z} \cT \right) c_l(\bk) \left( \cC_{2z} \cT \right)^{-1} = \sigma_x c_l(\bk)$ on the spinor $c_l(\bk) = (c_{lA}(\bk), c_{lB}(\bk))$, with $\sigma_x$ acting on the sublattice degree of freedom, as well as time-reversal $\cT$ which connects the two valleys. The combination of time-reversal $\cT$ and $\cC_{2z}$ symmetries fixes the phase structure of the numerically-obtained wavefunctions.

\section{Hartree corrections}
\label{app:Hartree}

We then consider the effects of Coulomb interactions, 
\begin{equation}
    H_C = \frac{1}{2} \int d^2\br d^2\br' \delta \rho(\br) V_C(\br - \br') \delta \rho(\br') ,
\end{equation}
with Coulomb potential $V_C(\br) = e^2/ (4 \pi \epsilon |\br| )$ and $\delta \rho (\br) = \sum_j \left( c_j^\dagger(\br) c_j (\br) - \langle c_j^\dagger(\br) c_j(\br) \rangle_{\nu = 0} \right)$ denotes the electronic density \emph{measured from the charge neutrality point, $\nu=0$.} Here $j$ is a combined spinor index that runs over both layer index $l$ and sublattice $A/B$. We treat the dielectric constant $\epsilon = \epsilon_r \epsilon_0$ as a free parameter to account phenomenologically for screening effects -- effectively adjusting the (dimensionless) relative permittivity $\epsilon_r$. Following Refs.~\citenum{guineaElectrostaticEffectsBand2018, ceaElectronicBandStructure2019, ceaBandStructureInsulating2020} we consider a mean-field decoupling of $H_C$ where we only keep the (local) Hartree correction
\begin{equation}
    H_H = \int d^2 \br V_H(\br) \sum_j c_j^\dagger(\br) c_j(\br),
\end{equation}
with the Hartree potential
\begin{equation}
    V_H(\br) = \int d^2 \br' V_C(\br - \br')  \sum_j \langle c_j^\dagger(\br') c_j(\br') \rangle_H,
\end{equation}
and the expectation value $\langle \hdots \rangle_H$ is taken relative to the charge neutrality point, $\nu=0$.

In the momentum-space basis defined by the Bloch wavefunctions of the BM model, The Hartree contribution reads
\begin{equation}
    \langle \bk + \bg, \tau, j | H_H | \bk + \bg' , \tau', j' \rangle = \delta_{j j'} \delta_{\tau \tau'} V_C(\bg - \bg') \delta \rho(\bg - \bg')
\end{equation}
with the Fourier-transformed Coulomb potential $V_c(\bg) = e^2/(2 \epsilon_r \epsilon_0 |\bg|)$, and
\begin{align}
\label{eq:app_hartree_self_energy}
    \delta \rho(\bg) = \sum_{\tau, n, \bm{k},\bg',j}' \left(U_{\tau n \bm{k}}^{j, \bg'} \right)^* U_{\tau n \bm{k}}^{j, \bg'-\bg}\,,
\end{align}
where $U_{\tau n \bm{k}}^{j,\bg}$ are the Bloch-wave expansion coefficients (see also Eq. \eqref{eq:Bloch_BM}) in valley $\tau$ and band $n$. In the above expression the $\sum'$ denotes summation over occupied states for a given filling measured with respect to the CNP as explained previously. Details of the self-consistent calculation were described in Refs.~\citenum{choiInteractiondrivenBandFlattening2021,kimEvidenceUnconventionalSuperconductivity2022,zhangPromotionSuperconductivityMagicangle2022}. Here $\delta\rho(\bg)$is the Fourier component of the charge density (relative to $\nu=0$), which is self-consistently determined by demanding that it leads to the same eigen-energies than those used to compute it. In practice~\cite{guineaElectrostaticEffectsBand2018,ceaElectronicBandStructure2019, ceaBandStructureInsulating2020}, it is sufficient to retain only the leading-order contributions in $V_C(\bg)$, i.e. the six terms with smallest momentum transfer, $\pm \bg_j$ with $j=1,2,3$. Because  $C_{3z}$ symmetry is broken due to heterostrain, both the Coulomb potential and the self-consistently determined $\delta \rho(\bg)$ parameters will also be direction dependent -- however time-reversal symmetry enforces that the $\pm$ contributions remain equal. Note that contributions with $\bg - \bg'=0$ are omitted, because they are cancelled by the ionic background.

\section{Strained bands and nesting instabilities to inter-valley coherent states}
\label{app:strained_bands}

\begin{figure}
    \includegraphics[width=\linewidth]{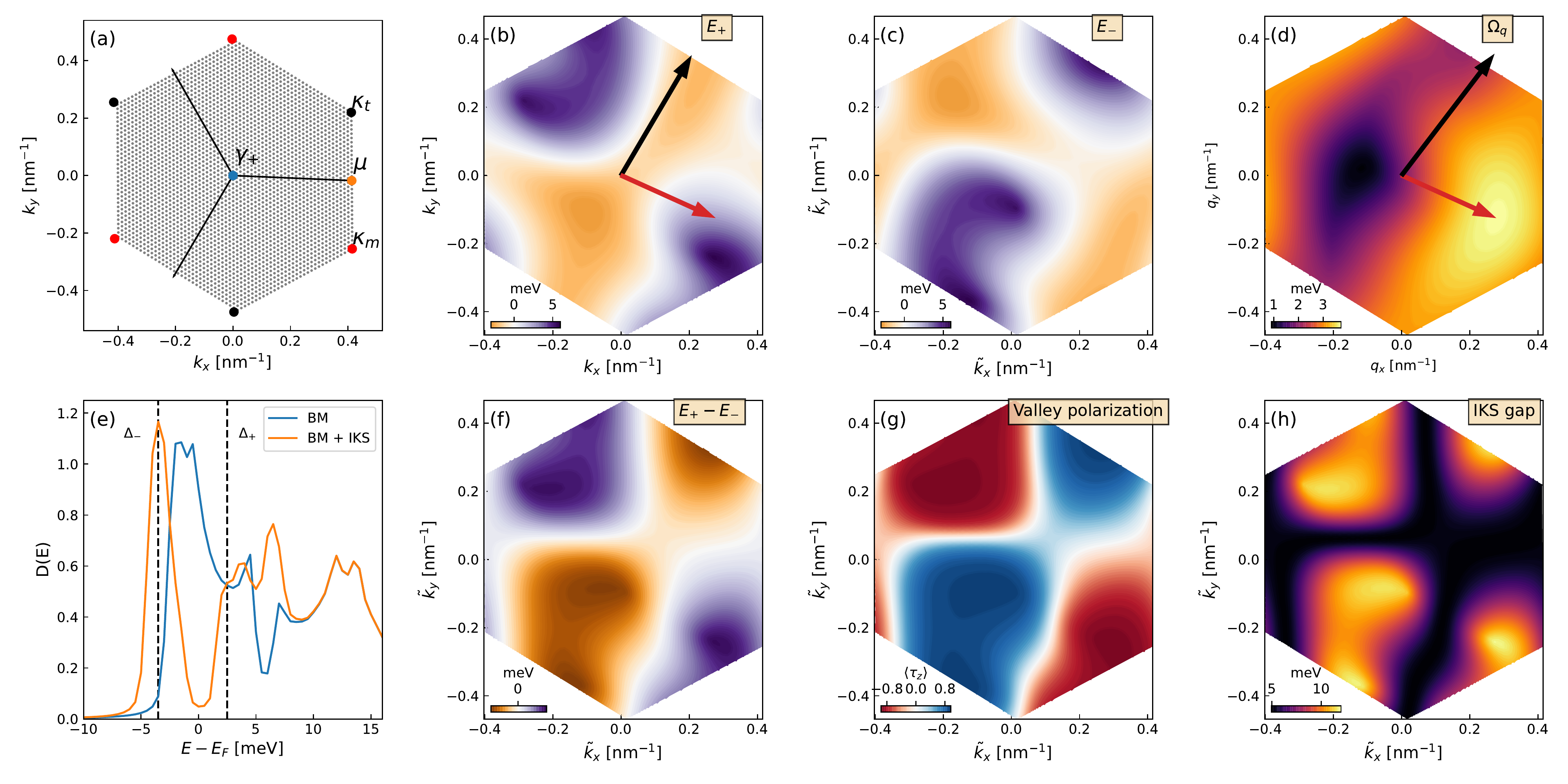}
	\caption{(a) Geometry of the strained moir\'e BZ, assuming strain of magnitude $\epsilon = -0.12 \%$ and direction $\varphi = 87^{\circ}$ and twist angle $\theta = 1.602^\circ$. The momenta $\bk$ are measured from the strained $\gamma_+$ point. Black and red circles denote Dirac points originating from the top/bottom and middle graphene layers, respectively. The three black arrows denote the direction of the strained reciprocal lattice vectors $\bg_j$. The gray dots show the momentum grid used for computations, including $n_k = 3072$ points.
    (b): The top-most valence band of the strained BM model in valley $\bK_+$, including self-consistent Hartree corrections with relative permittivity $\epsilon_r = 30$, with energies measured from the Fermi energy at $\nu=-2$. We use $(w_0, w_1) = (55, 105)$ meV and $v_F \approx 8.7 \times 10^5$ m/s, which yield a first magic magic angle $\theta_{m} \sim 1.52^\circ$. In the presence of strain the two Dirac cones are no longer tethered to their usual location at the corners of the moir\'e BZ due to the breaking of $C_3$ symmetry. We show the optimal IKS wavevector $\bq_{\rm IKS}$ by a red arrow, and the corresponding modulation wavevector $\bq_{\rm Kekul\acute{e}}$ by a black arrow. In panel (c) we show the $\bK_-$ valley (with identical parameters as in (b)), albeit shifted by $-\bq_{\rm IKS}$ -- i.e., plotting in terms of the renormalized momenta $\tilde{\bk} = \bk + (\tau_z -1) \frac{q_{\rm IKS}}{2}$. In the shifted coordinates the band maxima and minima of the two valleys are roughly aligned -- compare panels (b) and (c). (d) The optimal IKS wavevector is selected by maximizing the function $\Omega_{\bq}$ defined in Eq.~\ref{eq:Omega_q}, which computes the averaged energy separation between the valence bands in the two valleys when shifted by a relative wavevector $\bq$.
    (f) The energy difference $E_+(\tilde{\bk}) - E_-(\tilde{\bk})$ between the valence bands in the two valleys, in the shifted momentum coordinates, for the optimial $\bq_{\rm IKS}$. An IKS state takes advantage of the strained band structure by developing inter-valley coherence primarily for momenta $\tilde{\bk}$ where the two valleys are almost degenerate (white color in f), while keeping the other regions ``locally" valley polarized (see panel g). Panel (h) shows the momentum-dependent gap opened by an IKS order with strength $\Delta_{\rm IKS} = 2.5$ meV. (e) The density of states corresponding to panels (b) and (c) (blue line), compared with the addition of an IKS  order parameter with $\Delta_{\rm IKS} = 2.5$ meV (orange line). The IKS state opens a full gap at $\nu=-2$. The symbols $\Delta_{\pm}$ denote energies on either side of the IKS gap (black dashed lines) where local density of states calculations are reported in Fig.~\ref{fig:LDOS}.}
	\label{fig:IKS}
\end{figure}

Strained band structures are shown in Fig.~\ref{fig:IKS} (b),(c). As explained above, we fix the strain magnitude $\epsilon = -0.12 \%$ and angle $\varphi = 87 ^{\circ}$ to best reproduce the observed moir\'e Bravais lattice vectors in experiment (see also Fig.~\ref{fig:LDOS} (a).) We first consider the top-most valence band ot the BM model, with relatively weak electronic interactions ($\epsilon_r = 30$), as shown in Fig.~\ref{fig:IKS} (b) -- see also Fig~\ref{fig: fig4}j in the main text.
The flat bands are significant deformed by strain: in particular, the minimum remains in the region around the $\gamma$ point, but the maxima (located near the $\kappa$, $\kappa'$ points in pristine TTG) move significantly in both energy and momentum. In the presence of Coulomb interactions, Hartree corrections promote a band inversion mechanism rooted in the additional energy cost needed to host electronic states that have a strong spatial overlap\cite{guineaElectrostaticEffectsBand2018, ceaElectronicBandStructure2019, ceaBandStructureInsulating2020, rademakerChargeSmootheningBand2019, goodwinHartreeTheoryCalculations2020}. As shown in the main text Fig.~\ref{fig: fig4}k for $\epsilon_r = 15$, this results in a band where the maximum now occurs near $\gamma$ and the minimum occurs near one of the $\mu$ points.

The strained bands in both the weakly-interacting (non-inverted) and strongly-interacting (inverted) regimes are potentially susceptible to a type of ``nesting instability" between the two valleys, where the system develops  inter-valley coherence with a relative momentum shift $\bq_{\rm IKS}$ that roughly aligns the maximum of one valley with the minimum of the other, as shown in Fig.~\ref{fig:IKS} (b) and (c). This process allows a compromise between minimizing the exchange energy (from flavor polarization in the valley subspace) and a kinetic energy gain (from populating the lower-energy regions of the band structure in each valley~\cite{kwanKekulSpiralOrder2021, wagnerGlobalPhaseDiagram2022}). Motivated by such an energetic picture of the IKS order, we introduce a figure of merit for the optimal IKS wavevector as
\begin{equation}
    \Omega_{\bq} = \frac{1}{N} \sum_{\bk} \left| E_{+, \bk} - E_{-,\bk - \bq} \right| .
    \label{eq:Omega_q}
\end{equation}
This object computes (the absolute value of) the energy separation between the relevant bands in the two valleys $\tau=\pm$, averaged over a moir\'e Brillouin zone, when shifted by a relative momentum shift $\bq$. (Here $N$ is the number of moir\'e unit cells.) The optimal IKS wavevector $\bq_{\rm IKS}$ is then obtained by maximizing $\Omega_{\bq}$, as shown in Fig.~\ref{fig:IKS} (d). The energy separation between bands for the optimal $\bq_{\rm IKS}$ is also shown in Fig.~\ref{fig:IKS} (f).

Fig.~\ref{fig:IKS} (g) shows the valley polarization $\langle \tau_z \rangle$ in the IKS state constructed in such a way. The valley polarization is strongly momentum-dependent, and benefits from the ability to rotate within the Brillouin zone (in contrast to e.g. a T-IVC state). Further, developing inter-valley coherence with a finite momentum offset allows to efficiently open up a gap at $\nu=-2$, as shown in Fig.~\ref{fig: fig1} (e) and (h). Note that the T-IVC state can also open up a gap efficiently at $\nu=2$, by hybridizing the two Dirac points in the mini-BZ, provided that spin degeneracy is also spontaneously broken.

Comparing Fig.~\ref{fig: fig4} (j) and (k), it is clear that interaction-induced band inversion can have a dramatic effect on the leading IKS instability -- due to the discrete jump in the location of the band extrema as a function of interaction strength. In the weakly-interacting (non-inverted) regime, the preferred $\bq_{\rm IKS}$ connects points in the vinicity of $\gamma$ and $\kappa$, shifted and distorted by heterostrain. In contrast, in the strongly-interacting (inverted) regime, $\bq_{\rm IKS}$ rather connects points in the vicinity of $\gamma$ and $\mu$. Heterostrain breaks the degeneracy between the three $\mu$ points and selects the direction of the IKS instability.
In the inverted regime, strain competes against a larger (Hartree) energy scale, compared to the bare continuum model bandwidth, and is therefore less effective at distorting the band -- and as a consequence, modifying the magnitude and direction of the IKS wavevector. 

We study in more detail the physics of band inversion, and its influence on the IKS wavevector, as a function of the dielectric constant $\epsilon_r$ that sets the interaction energy scale. 
Our results are shown in Fig.~S\ref{fig:Hartree}. Generally we find that when the system lies in the band-inverted regime (small $\epsilon_{r}$), the direction of the IKS vector aligns with one of the $\gamma - \mu$ axes: for the strain parameters extracted from experiment, the direction of the $\bg_3$ moiré reciprocal vector is selected. Upon doping away from $\nu=-2$ (as Hartree effects increase and further drive the band inversion), the length of the IKS wavevector $\bq_{\rm IKS}$ slowly grows (see panel a). 
In the non-inverted regime obtained for weaker interactions (larger $\epsilon_r$), the magnitude of $\bq_{\rm IKS}$ is larger and varies more with doping, more in line with experimental results. Furthermore, its direction (shown in panel b) is consistent with the observed experimental direction, once accounting for the momentum shift between $\bq_{\rm IKS}$ and $\bq_{\rm Kekul\acute{e}}$ -- see also Fig.~\ref{fig: fig4} j and k.

Our analysis neglects the role of exchange interactions -- therefore, we do not determine the peferred IKS wavevector through a self-consistent solution but following a phenomenological procedure introduced in Ref.~\cite{kwanKekulSpiralOrder2021} and inspired by the energetics of the IKS state. Such approximations constitute limitations of our theory: specifically, a self-consistent procedure is necessary to capture both the feedback of the IKS order on the band dispersion, and also to determine whether an IKS state is the relevant ground state for a given set of BM model parameters.
Exclusion of the exchange (Fock) effects and displacement field $D$ also neglects the broadening of the MATTG flat bands\cite{xieWeakFieldHallResistivity2021,xieNatureCorrelatedInsulator2020,ceaBandStructureInsulating2020}, which counteract to some extent Hartree effects\cite{choiInteractiondrivenBandFlattening2021} and thus renormalizes the parameters required for the band-inversion transition. A separate self-consistent study of the influence of all these parameters on the IKS ground state in MATTG is however called for.

\section{Modeling of inter-valley coherent orders}
\label{app:modeling_IVC}

We now discuss in more detail symmetry-breaking orders that give rise to lattice tripling, focusing on IKS states which were found to be most consistent with the experimental data. We also briefly comment on moir\'e-periodic IVC states as a useful reference point.

We adopt the band basis of the (Hartree-renormalized) BM model. Such a basis is defined by operators $f_{\tau n}(\bk)$, which are related to the original (graphene) operators $c_j$ used in the previous section by diagonalizing the BM model
\begin{equation}
    f^\dagger_{\tau n}(\bk) = \sum_{j \bg} U^{j \bg}_{\tau n \bk} c^\dagger_j(\gamma_\tau + \bk + \bg),
\end{equation}
In the following we focus on the active band of interest: at $\nu=-2$, the top-most valence band of the strained BM model. We drop the band index $n$ and denote simply $f(\bk) \equiv \left( f_{+}(\bk), f_{-}(\bk) \right)^T$ the operators that annihilate electrons in valleys $\pm$ and mBZ momentum $\bk$ in this band. 
In this basis, we can write an ansatz for a moir\'e-periodic IVC state as
\begin{align}
H_{\rm TIVC} = \sum_{\bk}f^\dagger(\bk) \Delta_{\rm IVC}(\bk)   f(\bk) ,
\label{eq:TIVC}
\end{align}
with $\Delta_{\rm IVC}(\bk)$ an off-diagonal matrix acting in valley space, such that $\tau_x \Delta^*(-\bk) \tau_x = \Delta(\bk)$ in order to respect time-reversal symmetry. We note that this ansatz differs from the T-IVC state proposed in strong-coupling approaches, where the development of inter-valley coherence gaps out the Dirac cones near the $\kappa$ and $\kappa'$ points in the moir\'e BZ. At $\nu=-2$, for such a mechanism to be operative the spin degeneracy of the flat bands in each valley must first be lifted -- either leading to a fully spin-polarized or a spin-valley-locked insulating state. 

In contrast, IKS states are parametrized by a (generally incommensurate) wavevector offset $\bq_{\rm IKS}$ between the two valleys. The IKS order parameter is best described by defining a valley-dependent shifted momentum as $\tilde{\bk} = \bk + (\tau-1) \bq_{\rm IKS}/2$ -- this is allowed because of a new Bloch theorem associated with an effective translation symmetry~\cite{kwanKekulSpiralOrder2021}. We can organize the electron operators above as $f(\tilde{\bk}) \equiv  (f_{+}(\tilde{\bk}), f_{-}(\tilde{\bk}))^T = \left( f_{+}(\bk), f_{-}(\bk - \bq_{\rm IKS}) \right)^T$. In this basis, one can write an ansatz for the IKS state as 
\begin{align}
H_{\rm IKS} = \sum_{\tilde{\bk}} f^\dagger(\tilde{\bk}) \Delta_{\rm IKS}(\tilde{\bk}) f(\tilde{\bk}) .
\label{eq:IKS_shifted}
\end{align}
The intuition behind this uniform shift is that it allows the maximum of one valley to lie on top of the minimum of the other valley. Then, an inter-valley type mass term can efficiently open a gap near $\nu=\pm 2$ whilst allowing the system to remain valley-polarized for momenta where that is energetically favorable -- thus gaining some kinetic energy, at the expense of an exchange energy cost compared to a uniformly polarized IVC state.

For numerical simulations, we take the following ansatze for the inver-valley coherent order parameters. For the moir\'e-periodic IVC order we take a simple momentum-independent form 
$\Delta_{\rm IVC}(\bk) = \Delta_{\rm IVC} \tau_x$.
For the IKS state, inspired by self-consistent numerical treatments~\cite{kwanKekulSpiralOrder2021, wagnerGlobalPhaseDiagram2022, wangKekulSpiralOrder2022}, we take an ansatz where the order parameter ``locally" adjusts
to the band structure in the shifted coordinates $\tilde{\bk}$ -- see Fig.~\ref{fig:IKS} f. The physical intuition behind the energetics of the IKS state is that valley polarization is preferred when the difference in energy $\Delta E(\tilde{\bk}) = E_+(\tilde{\bk}) - E_-(\tilde{\bk})$ in the two valleys is large, while inter-valley coherence is favored when the difference is small. We therefore take an ansatz
\begin{equation}
\Delta_{\rm IKS}(\tilde{\bk}) = \Delta_{\rm IKS} \left( \cos( \theta^{\rm IKS}_{\tilde{\bk}} ) \tau_x + \sin( \theta^{\rm IKS}_{\tilde{\bk}} ) \tau_z \right) ,
\end{equation}
where $\theta^{\rm IKS}_{\tilde{\bk}} = \arctan \left( \Delta E(\tilde{\bk})/\gamma \right)$ parameterizes the polar angle (measured from the equator) of the valley pseudospin on the Bloch sphere. We take the IKS order parameter amplitude $\Delta_{\rm IKS} = 2.5$ meV to roughly match the energy scale of the spectroscopic (pseudo-)gap (see Fig.~\ref{exfig: Kekule_bias} b) and the ``tilt parameter" $\gamma = 10$ meV to prepare plots in Figs.~\ref{fig:IKS} and \ref{fig:LDOS}.

\section{Real-space LDOS features}
\label{app:LDOS_method}

\begin{figure}
	\includegraphics[width=\linewidth]{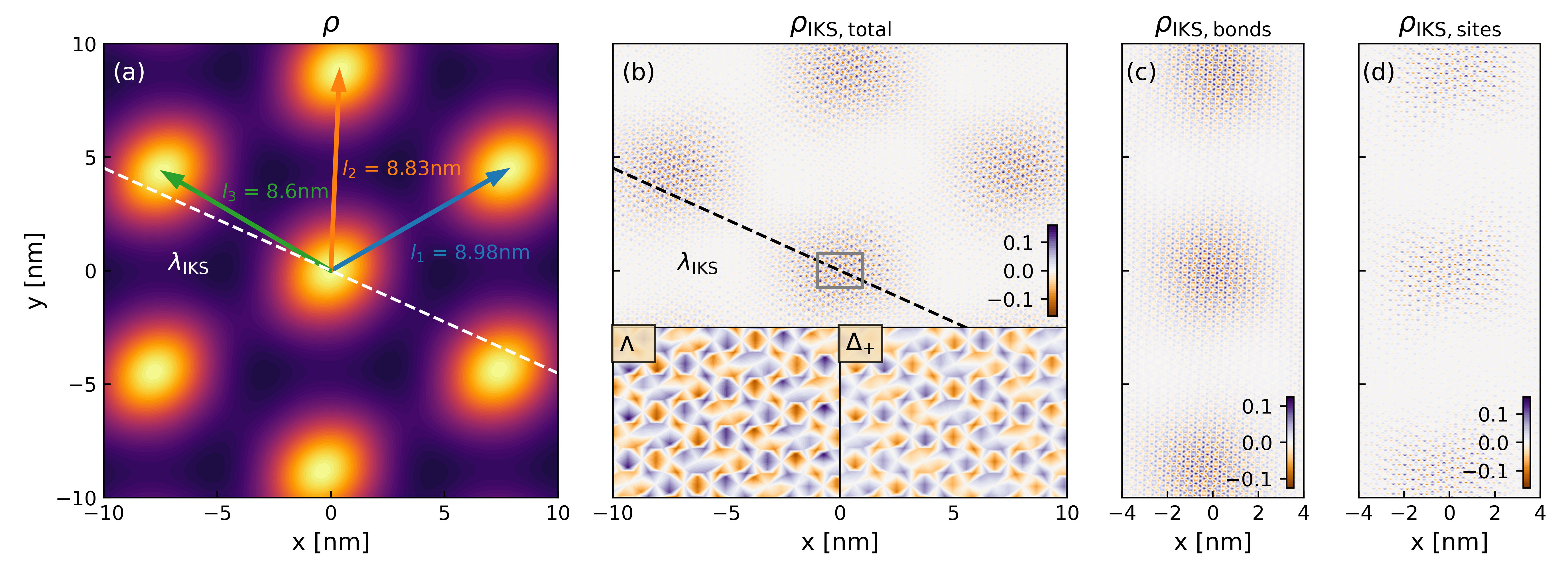}
   	\includegraphics[width=\linewidth]{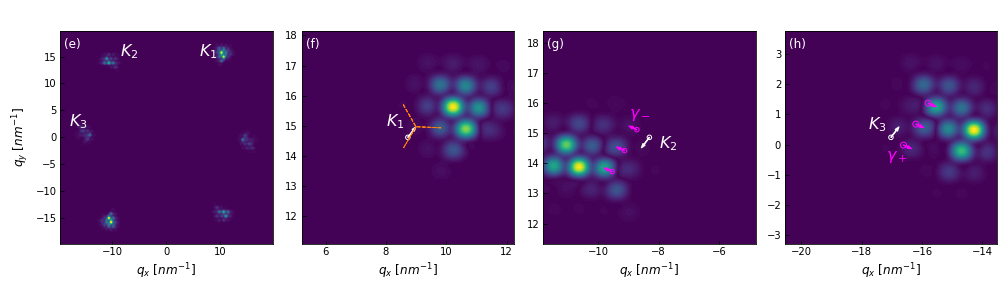} 
	\caption{Signatures of IKS order in local density of states $\rho(\br, E)$, using the same parameters as for Fig.~\ref{fig:IKS} (g)-(h). Panel (a) shows the continuum charge density profile computed from Eq.~\ref{eq:rho_eigenstate},  at energy $E = \Delta_-$ (at the DOS peak below the IKS gap). The charge density is peaked at moir\'e AAA sites. The three Bravais lattice vectors have lengths $l_1$, $l_2$ and $l_3$ that match the experiment. We also note the direction of the IKS modulation by a white dashed line. Panel (b) shows the total lattice-tripling signal at $E=\Delta_-$. The two insets show a close-up of the grey box, for $\Delta_-$ as well as $\Delta_+$ (just above the IKS gap). The contrast is inverted between the two images. Panels (c) and (d) show the site-resolved (Eq.~\ref{eq:IVC_sites}) and bond-resolved (Eq.~\ref{eq:IVC_bonds}) channels along a stripe. Here we use ``half visibility" for the bond vs site orders, $\Phi_1 = \Phi_0/2$. The incommensurability of the lattice-tripling pattern with the moire unit cell is visible by comparing neighboring AAA sites. Panels (e) to (h) show the Fourier-transformed lattice-tripling signal (from panel b) at $E=\Delta_-$. Panel (e) shows all six (first-order) lattice tripling peaks near momenta corresponding to the graphene Brillouin zone corners. A zoomed-in view of the three inequivalent regions near $\bK_1$, $\bK_2$, $\bK_3$ is shown in panels (f) to (h). Satellite peaks separated by (strained) moir\'e reciprocal lattice vectors $\bg$ (denoted by dashed orange lines in (f)) are resolved due to the large simulated real-space area that include multiple moir\'e unit cells. The peaks are however shifted from the expectation for moir\'e-translation invariant states
    (the magenta circles denote different mini-BZ centers $\gamma_\tau$) by the incommensurate modulation $\bq_{\rm IKS}$ (magenta arrow). Equivalently, the modulation is shifted from the graphene top-layer zone corners $\bK_1$, $\bK_2$ and $\bK_3$ (white circles) by the momentum shift $\bq_{\rm Kekul\acute{e}}$ (white arrow) defined in the main text (see also Fig.~\ref{fig: fig4} h). The momentum shift is opposite near $\bK_2$, as expected from time-reversal symmetry.}
	\label{fig:LDOS}
\end{figure}

\begin{figure}
    \includegraphics[width=\linewidth]{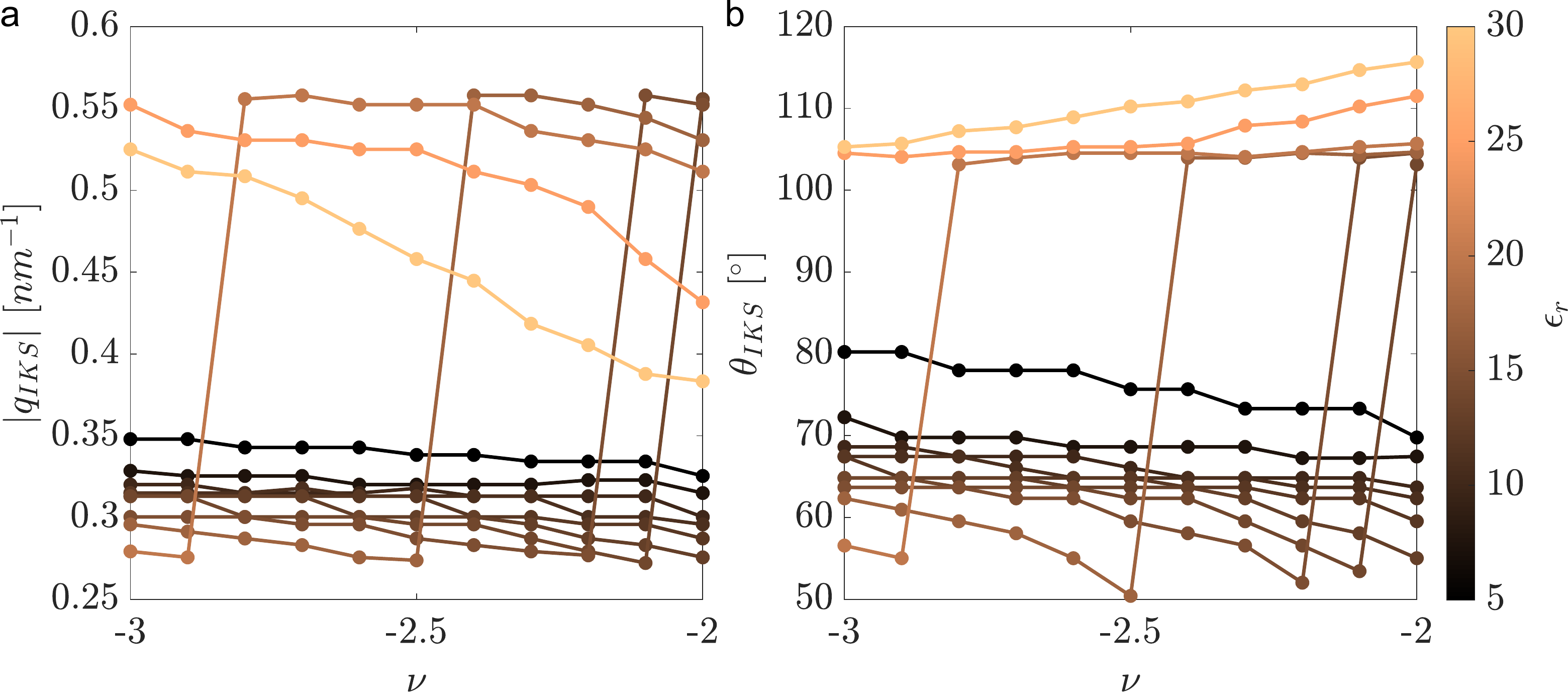}
 \caption{Hartree corrections and their consequences for IKS instabilities. All results are obtained from a self-consistent calculation of Hartree effects in MATTG with the IKS wavevector $\bq_{\rm IKS}$ determined through a procedure of aligning maxima and minima of bands in different valleys, inspired by Ref. \citenum{kwanKekulSpiralOrder2021}. The parameters used are $\theta = 1.602 ^\circ$, $w_0 = 55$ meV, $w_1 = 105$ meV, $v_F = 8.7 \times 10^{5}$ m/s
 and strain parameters $\epsilon_{\rm str} = -0.12 \%$, $\varphi = 87^\circ$. The magnitude (panel a) and direction (panel b) of the preferred IKS wavevector $\bq_{\rm IKS}$ are plotted as a function of filling fraction $\nu$ between $-3$ and $-2$ and $\epsilon_r$.
 }
	\label{fig:Hartree}
\end{figure}

The Bloch wavefunctions corresponding to the eigenstates of the BM model read
\begin{equation}
    \psi^j_{\tau n \bk}(\br) = e^{i (\gamma_\tau + \bk) \cdot \br} \sum_{\bg} e^{i \bg \cdot \br} U_{\tau n \bk}^{j \bg}
    \label{eq:Bloch_BM} ,
\end{equation}
The set of vectors $\bg = n_1 \bg_1 + n_2 \bg_2$ with $n_1, n_2$ integers are moir\'e reciprocal lattice vectors, truncated at finite values $-2 < n_i < 2$ in practice, and $\gamma_\tau + \bk$ is the microscopic momentum associated with the electronic state labeled by the moir\'e mBZ momentum $\bk$. The coefficients $U_{\tau n \bk}^{j \bg}$ are obtained from diagonalizing the BM model -- including Hartree corrections, as described in Sec.~\ref{app:Hartree} -- and correspond to eigenvalues $E_{\tau n \bk}$ labeled by the valley $\tau$, band index $n$ and mBZ momentum $\bk$.
The continuum (coarse-grained) charge density corresponding to each eigenstate can be expressed as
\begin{equation}
    \rho_{\tau n \bk}(\br) = \sum_j |\psi^j_{\tau n \bk}(\br)|^2 =  \sum_{j \bg \bg'} e^{i (\bg - \bg') \cdot \br} U_{\tau n \bk}^{j \bg} \left( U_{\tau n \bk}^{j \bg'} \right)^*.
    \label{eq:rho_eigenstate}
\end{equation}
From this expression the local charge density at energy $\omega$ and position $\br$ can be computed using
\begin{equation}
    \rho(\br, \omega) = -\frac{1}{\pi} \sum_{\tau n \bk} \text{Im} \left[\frac{ \rho_{\tau n \bk}(\br) }{\omega + i \eta - E_{\tau n \bk}} \right],
    \label{eq:LDOS}
\end{equation}
with small $\eta = 0.2$ meV to simulate thermal broadening corresponding to $T \sim 2$ K. A typical charge density profile in the moire unit cell is shown in Fig.~\ref{fig:LDOS}a, with strain parameters relevant to the experiment and self-consistent Hartree corrections with dielectric constant $\epsilon_r=30$.

In the presence of electronic interactions, the system may favor symmetry breaking in the spin-valley subspace. In particular, inter-valley coherent orders, which couples the two valleys, leads to eigenvalues $E_{m \bk}$ associated with linear combinations of valley eigenstates of the form
\begin{align}
    \Psi^j_{m \bk}(\br) &= \sum_{\tau n} C^{\tau n}_{m \bk} \psi^j_{\tau n \bk}(\br) = \sum_{\bg \tau n} e^{i (\gamma_\tau + \bg + \bk) \cdot \br}  C^{\tau n}_{m \bk} U_{\tau n \bk}^{j \bg}.
    \label{eq:Psi_R}
\end{align}
The coefficients $C^{\tau n}_{m \bk}$ characterizing the inter-valley coherence can be obtained, e.g., from fully self-consistent Hartree-Fock ~\cite{kwanKekulSpiralOrder2021, wagnerGlobalPhaseDiagram2022} or tensor-network~\cite{wangKekulSpiralOrder2022} calculations. In this work, we adopt a more phenomenological approach and obtain the $C^{\tau n}_{m \bk}$ in two steps: First, we consider self-consistent Hartree corrections to the band structure obtained from the BM model, and then we re-diagonalizing the effective Hamiltonian with inter-valley coherent mass terms, Eq.~\ref{eq:TIVC} or ~\ref{eq:IKS_shifted} depending on the type of order we wish to simulate. Motivated by modeling the STM data near $\nu=-2$, we retain only the top-most valence band of the BM model; the band index $n$ will therefore be dropped from here on.
Here we work in the basis of (Hartree-renormalized) BM bands, rather than in the Chern basis used in Ref.~\cite{hongDetectingSymmetryBreaking2022, calugaruSpectroscopyTwistedBilayer2022}.
In general, the BM bands consist of a linear combination of the Chern basis eigenstates -- this will have consequences later on when analyzing the LDOS patterns in real space. We also note that when considering IKS order parameters, the momentum that labels the eigenstates will be $\tilde{\bk}$ instead of $\bk$: see also the comment after Eq.~\ref{eq:LDOS_channels}.

The (continuum) charge density corresponding to a state with inter-valley coherence reads
\begin{widetext}
\begin{equation}
    \rho_{m \bk}(\br) = \sum_j | \Psi^j_{m \bk}(\br) |^2
    = \sum_{j, \bg, \bg', \tau, \tau'} e^{i (\gamma_{\tau} - \gamma_{\tau '} + \bg - \bg') \cdot \br} C^{\tau }_{m \bk}  \left(C^{\tau' }_{m \bk} \right)^* U_{\tau  \bk}^{j \bg} \left( U_{\tau'  \bk}^{j \bg'} \right)^*.
\end{equation}
\end{widetext}
However, this quantity cannot capture Kekul\'e bond orders because it is a sum of local densities at each sublattice site $j = (A_1, B_1, A_2, B_2, A_3, B_3)$. Therefore, in the next section we develop a scheme, partially inspired by Ref.~\cite{hongDetectingSymmetryBreaking2022, calugaruSpectroscopyTwistedBilayer2022}, to efficiently evaluate IVC order parameters on both lattice sites and bonds, in large real-space areas containing multiple moir\'e unit cells, comparable to those obtained in the experiment.

\subsection{IVC order and lattice-tripling patterns}

To analyze inter-valley coherent orders in real-space LDOS maps, we need to consider both on-site charge densities as well as bond orders. The on-site charge density signal arises from intra-sublattice contributions, whereas bond orders arise from inter-sublattice terms. Going beyond the continuum description (where $\br$ is a continuum variable), we need to remember that charge physically originates from the $p_z$ atomic orbitals that are centered on discrete lattice sites $\br_j$.

We thus need to revisit the expansion in the BM model that expresses its eigenfunctions in terms of plane waves. In other words, we need to consider terms of the form

\begin{equation}
    \langle \br | c_{\tau j}^\dagger(\gamma_\tau + \bk + \bg) | 0 \rangle ,
\end{equation}
where $c^\dagger_j$ above creates an electron in the low-energy expansion of monolayer graphene, with spinor index $j = (\alpha, l)$ and microscopic momentum $\gamma_\tau + \bk + \bg$. In the ``coarse-grained" or continuum approach, one simply takes the plane-wave ansatz where
\begin{equation}
    \langle \br | c_j^\dagger(\gamma_\tau + \bk + \bg) | 0 \rangle = e^{i (\gamma_\tau + \bk + \bg) \cdot \br} ,
    \label{eq:Psi_R_planewaves}
\end{equation}
which leads to the Bloch wavefunction in Eq.~\ref{eq:Bloch_BM}. Incorporating the lattice, one can Fourier transform using
\begin{align}
    c^\dagger_j(\bk) = \sum_{\br_j} e^{i \bk \cdot \br_j } d^\dagger(\br_j),
\end{align}
where $\br_j = \bR_j + \tau_j$. Here $\bR_j$ is a Bravais lattice vector that labels the unit cell, and $\tau_j$ the location of the orbital on sublattice $j$ within that unit cell. The operator $d^\dagger( \br_j )$ creates an electron in a $p_z$ orbital at site $\br_j$, i.e. 
\begin{equation}
    \langle \br | d^\dagger (\br_j) | 0 \rangle = \Phi(\br - \br_j),
\end{equation}
where the electronic wavefunction around lattice site $\br_j$ can be approximate by a Gaussian orbital,
\begin{equation}
\Phi(\br - \br_j) = \frac{1}{\sqrt{2 \pi} \sigma} e^{-\frac{(\br - \br_j)^2}{2 \sigma^2}}
\end{equation}
with standard deviation $\sigma$. We then have
\begin{equation}
    \langle \br | c^\dagger_{j}(\gamma_\tau + \bk + \bg) | 0 \rangle = \sum_{\br_j} e^{i (\gamma_\tau + \bk + \bg) \cdot \br_j }  \Phi( \br - \br_j )
\end{equation}
and thus the Bloch wavefunction associated with band $m$ and wavevector $\bk$ takes the form
\begin{align}
    \Psi_{m \bk}(\br) 
    &= \sum_{\bg, \tau, j, \br_j} C^{\tau }_{m \bk} U_{\tau  \bk}^{j \bg} e^{i (\gamma_\tau + \bk + \bg) \cdot \br_j }  \Phi( \br - \br_j )
    \label{eq:Psi_R_discrete}
\end{align}
To recover the continuum description, one can take the limit $\sigma \rightarrow 0$ to make the orbitals point like, $\Phi(\br) \sim \delta(\br)$, in which case Eq.~\ref{eq:Psi_R_discrete} reduces to Eq.~\ref{eq:Psi_R}.

\subsection{Inter-valley coherence: site vs bond order}

Consider the top layer components of the wavefunction in Eq.~\ref{eq:Psi_R_discrete} (as the STM tip primarily couples to it). The sublattice indices $j$ now become $\alpha, \beta=A,B$. We denote by $\br_\alpha = \bR_\alpha + \tau_\alpha$ the positions of all the orbitals (or graphene sites) on sublattice $\alpha$. The charge density $\rho_{m \bk}(\br)  = |\Psi_{m \bk}(\br)|^2$ for each mode $m \bk$ reads
\begin{align}
    \rho_{m \bk}(\br) 
    &= \sum_{\substack{\bg, \bg', \tau, \tau', \\ \alpha, \beta, \br_\alpha, \br_\beta}} F^{\tau \alpha \bg}_{m \bk}  \left(  F^{\tau' \beta \bg'}_{m \bk} \right)^* e^{i (\gamma_\tau + \bk + \bg) \cdot  \br_\alpha }  e^{-i (\gamma_{\tau'} + \bk + \bg') \cdot \br_{\beta} } \Phi( \br - \br_\alpha)  \Phi^*( \br - \br_\beta) ,
    \label{eq:rho_R_discrete}
\end{align}
where for simplicity of notation we defined $F^{\tau \alpha \bg}_{m \bk} = C^{\tau}_{m \bk} U_{\tau \bk}^{\alpha \bg}$.
Computing this object for a generic grid for $\br$ is computationally intensive. However, we can simplify this expression by keeping only two types of term. First, terms with $\alpha=\beta$ correspond to on-site charge densities. In this case we keep only $\br_\alpha = \br_\beta$ contributions and neglect contributions from neighboring orbitals. The on-site contribution therefore reads
\begin{align}
    \rho^{\rm site}_{m \bk}(\br) 
    &= \sum_{\bg, \bg', \tau, \tau' \alpha, \br_\alpha}  F^{\tau \alpha \bg}_{m \bk}  \left(  F^{\tau' \alpha \bg'}_{m \bk} \right)^* e^{i (\gamma_\tau + \bg - \gamma_\tau' - \bg') \cdot \br_{\alpha}} |\Phi( \br - \br_\alpha)|^2
    \label{eq:rho_R}
\end{align}
We can then sample $\br$ only on the actual graphene lattice sites, replacing the continuum variable $\br \rightarrow \br_\alpha$. The intra-valley ($\tau=\tau'$) contribution reads
\begin{align}
    \rho^{\rm site}_{m \bk}(\br_\alpha) 
    &= |\Phi_0|^2 \sum_{\bg, \bg', \tau} F^{\tau \alpha \bg}_{m \bk}  \left(  F^{\tau \alpha \bg'}_{m \bk} \right)^* e^{i (\bg - \bg') \cdot \br_\alpha}
    \label{eq:rho_R_discrete_diagonal}
\end{align}
where $|\Phi_0|^2$ is a number characterizing the on-site orbital visibility (which is a function of the Gaussian orbital width $\sigma$.) This expression is in essence a discretized version of the continuum charge density in Eq.~\ref{eq:rho_eigenstate}. We can also have a site-centered signal that breaks the translation symmetry of the graphene lattice. Such a contribution comes from inter-valley ($\tau \neq \tau'$) components of Eq.~\ref{eq:rho_R} and reads
\begin{align}
    \rho^{\rm site - IVC}_{m \bk}(\br_\alpha) 
    &= |\Phi_0|^2 \sum_{\tau} e^{2 i \gamma_\tau \cdot \br_{\alpha}} \sum_{\bg, \bg'} F^{\tau \alpha \bg}_{m \bk}  \left(  F^{\overline{\tau} \alpha \bg'}_{m \bk} \right)^* e^{i (\bg - \bg') \cdot \br_{\alpha}}
    \label{eq:IVC_sites}
\end{align}
with $\overline{\tau}$ denotes the opposite valley of $\tau$, and we used that $\gamma_{\overline{\tau}} = - \gamma_\tau$.

For the lattice-tripling bond order, we consider both valley and sublattice off-diagonal contributions, $\tau \neq \tau'$ and $\alpha \neq \beta$ in Eq.~\ref{eq:rho_R_discrete}. We want to evaluate this object on all the \emph{nearest-neighbor bonds}, i.e. for $\br \equiv \br_{\rm AB} = (\br_{\rm A} + \br_{\rm B})/2$. Let us also define $\delta \br_{\rm AB} \equiv (\br_{\rm A} - \br_{\rm B})/2$, the three different vectors than connect a particular site $B$ to its three closest bond centers. Throwing away all contributions from orbitals beyond nearest neighbors, we get
\begin{align}
    \rho^{\rm kekule}_{m \bk}(\br_{\rm AB}) 
    =& \sum_{\bg \bg' \tau} F^{\tau A \bg}_{m \bk}  \left(F^{\overline{\tau} B \bg'}_{m \bk} \right)^* e^{i (\gamma_\tau + \bk + \bg) \cdot \left( \br_{\rm AB} + \delta \br_{\rm AB} \right)}  e^{-i (\gamma_{\overline{\tau}} + \bk + \bg') \cdot \left( \br_{\rm AB} - \delta \br_{\rm AB} \right) } \Phi(-\delta \br_{\rm AB})  \Phi^*( \delta \br_{\rm AB} ) \nonumber \\
    &~~~~ + ( A \leftrightarrow B) \nonumber \\
    =& |\Phi_1|^2 \sum_{\bg \bg' \tau} F_{m  \bk}^{\tau A \bg} \left( F_{m \bk}^{\overline{\tau} B \bg'} \right)^*  e^{i (2 \gamma_\tau + \bg - \bg') \cdot \br_{\rm AB} }  e^{i (\bg + \bg' + 2 \bk) \cdot \delta \br_{\rm AB}} + ( A \leftrightarrow B) , 
    \label{eq:IVC_bonds}
\end{align}
where in the second line we assumed that the overlap between atomic orbitals on the three bonds, $\Phi(-\delta \br_{\rm AB})  \Phi^*( \delta \br_{\rm AB})$, is independent of the direction of the bond and just a number $|\Phi_1|^2$.

We therefore reduced the problem to computing the local density of states on two interwoven grids: one for the sites $\br_\alpha$ and one for the bonds $\br_{\rm AB}$. This represents a numerical speed-up by a few orders of magnitude over a brute-force approach. From these expressions we can compute the local density of states (LDOS) at energy $\omega$ and position $\br$, similarly as before, in various channels labeled by $c =$ (site, site-IVC, Kek).
\begin{equation}
    \rho^c(\br, \omega) = -\frac{1}{\pi} \sum_{m \bk} \text{Im} \left[\frac{ \rho^c_{m \bk}(\br)}{\omega + i \eta - E_{m \bk}} \right] .
    \label{eq:LDOS_channels}
\end{equation}

The expressions derived above work as stated for moir\'e-periodic inter-valley coherent states. However, for IKS states we must  remember that the momentum label for the eigenstates is actually $\tilde{\bk}$. However, the physical momentum appearing in the Fourier transforms must remain $\bk$, which is related to $\tilde{\bk}$ by $\bk = \tilde{\bk} - (\tau-1) \frac{\bq_{\rm IKS}}{2}$. We can therefore use the above expressions (Eqs.~\ref{eq:rho_R_discrete_diagonal}, \ref{eq:IVC_sites} and \ref{eq:IVC_bonds}), but where the sum over momentum labels runs over $\tilde{\bk}$, and the mBZ centers are appropriately shifted, $\gamma_\tau \rightarrow \gamma_\tau - (\tau-1) \frac{\bq_{\rm IKS}}{2}$.

\section{Analysis of LDOS patterns for inter-valley coherent states}
\label{app:LDOS_results}

Using the above approach we compute the lattice-tripling LDOS signal, using both the site-resolved and bond-resolved expressions Eqs.~\ref{eq:IVC_sites} and \ref{eq:IVC_bonds}. For both moir\'e-periodic IVC (not shown here) and IKS states (shown in Fig.~\ref{fig:LDOS}(b)--(d)), the lattice-tripling pattern is strongest around the AAA regions of the moir\'e unit cell, and its contrast is reversed when comparing energies just above and below the Fermi energy -- two features that are also present in the experimental data. The main difference between the two candidate orders is that in an IKS state, the symmetry-breaking pattern winds along a real-space direction set by $\bq_{\rm IKS}$. The corresponding modulation wavelength $\lambda_{\rm IKS} = 2 \pi / |\bq_{\rm IKS}|$ is generally incommensurate with the size of the moir\'e unit cell, such that the real-space symmetry-breaking pattern differs between neighboring AAA sites~\cite{kwanKekulSpiralOrder2021} (except in the direction perpendicular to $\bq_{\rm IKS}$). The relative magnitude of the bond-centered and the site-centered channels depends on the ratio $\Phi_1/\Phi_0$ which characterizes the width of the $p_z$ orbitals: we take a ratio $\Phi_1/\Phi_0 = 1/2$ in the plots in Fig.~\ref{fig:LDOS}, which leads to a substantial bond signal.

We stress that our analysis is based on instabilities of the (Hartree-renormalized) bands of the BM model, which consist of a linear combination of different Chern sectors. In the chiral limit the Chern sectors are defined by the relation $\cC = \tau_z \sigma_z$, i.e. they comprise states that are valley-sublattice locked. Therefore, IVC order parameters can be conveniently decomposed into intra-Chern and inter-Chern components. The resulting density of states  respectively live on the bonds -- for intra-Chern components, which couple e.g. $(\bK, A)$ and $(\bK', B)$ -- and sites -- for inter-Chern components, which couple e.g. $(\bK, A)$ and $(\bK',A)$ -- of the graphene lattice.  Therefore, in our case both site-and bond-centered signals are generically expected, and indeed observed in Fig.~\ref{fig:LDOS}.

The key difference between the LDOS of the IKS and moir\'e-periodic IVC states resides in the breaking of translation symmetry on the moir\'e scale. Such a feature can be most cleanly captured by considering the Fourier-transformed lattice-tripling LDOS signal,
\begin{align}
    \rho^{\rm lt}(\bq, \omega) = \sum_{\br} e^{i \bq \cdot \br} W(\br) \rho^{\rm lt}(\br, \omega) ,
    \label{eq:LDOS_FT}
\end{align}
where we defined $\rho^{\rm lt}(\br, \omega) \equiv \rho^{\rm Kek}(\br, \omega) + \rho^{\rm site-IVC}(\br, \omega)$, and to reduce finite-size effects we introduced the Hanning window
\begin{equation}
    W \left( \br = (x, y) \right) = \cos^2 \left( \frac{\pi y}{2 L_y} \right) \cos^2 \left( \frac{\pi x}{2 L_x} \right),
\end{equation}
where $L_{x,y}$ are the lengthscales that define the field of view along the $x$ and $y$ directions.

The calculated Fourier-transformed lattice-tripling signals are shown in Fig.~\ref{fig:LDOS} (e)-(h). The lattice-tripling signal occurs, as expected, near momentum transfer $\bK_1$, $\bK_2$ and $\bK_3$ that correspond to the corners of the Brillouin zone of monolayer graphene (panel e). A zoomed-in view near each of those momenta however reveals a much richer structure. First, a number of ``satellite" peaks, translated by integers of the (strained) moir\'e reciprocal lattice vectors $\bg$, captures the intra-unit cell structure of the inter-valley coherence order. Such peaks are however shifted from naive expectations: none of the peaks line up with the moir\'e BZ centers $\gamma_\pm$, denoted by magenta circles in each of the panels Fig.~\ref{fig:LDOS} (f)-(h), which characterize the LDOS in a moir\'e-periodic state such as Eq.~\ref{eq:TIVC}. Instead, the peaks are shifted by the IKS wavevector $\bq_{\rm IKS}$. Equivalenty, the observed peaks are shifted by the momentum $\bq_{\rm Kekul\acute{e}}$ from the position of the strained Dirac points $\bK_1$, $\bK_2$, $\bK_3$ as shown by the white circles and arrows in Fig.~\ref{fig:LDOS} (f)-(h) (see also Fig.~\ref{fig: fig4} h for a schematic of the modulation wavevector extraction from FT LDOS maps).

We also observe that the IKS state breaks $C_3$ symmetry on the microscopic graphene scale, as evident from the different Fourier-transformed intensities near the Brillouin zone corners $\bK_1$, $\bK_2$ and $\bK_3$, see panel (e). This rotation-symmetry breaking feature seems to be also observed in the experiment (compare to Fig.~\ref{fig: fig4}, panels e to g). We also observe more subtle features in the variation of intensity between satellite peaks, which seems to be cut off abruptly along one-dimensional lines (e.g. near $\bK_1$, panels e and f). This feature is similar to the ``sashes" identified theoretically in Ref.~\cite{hongDetectingSymmetryBreaking2022} and highlighted in Fig.~\ref{fig: fig4} a.

\section{Theoretical scenario for an incommensurate-commensurate transition in the Kekul\'e spiral}
\label{app:comensurate}

In this Section we investigate possible commensuration effects on the IKS states, motivated by the data in Fig.~\ref{fig: fig3}. We first review the Ginzburg-Landau analysis of IKS order parameters in Ref.~\citenum{kwanKekulSpiralOrder2021} and comment on new terms arising from the breaking of $U(1)_V$ symmetry by the lattice.

The Kekul\'e spiral order parameter can be expressed as a vector $\bI_{\bq} \equiv \left( I_{\bq}^x, I_{\bq}^y \right)$ with
\begin{align}
    I^x_{\bq} &= \frac{1}{N} \sum_{\bk} \langle \psi^\dagger_{\bk + \bq} \tau_x \psi_{\bk} \rangle ~ , &
    I^y_{\bq} &= \frac{1}{N} \sum_{\bk} \langle \psi^\dagger_{\bk + \bq} \tau_y \psi_{\bk} \rangle,
\end{align}
where $\psi_{\bk}^\dagger = \left( \psi_{\bk, +}^\dagger, \psi_{\bk, -}^\dagger \right)$ denote creation operators for electrons in the active band of interest (e.g., the top-most valence band at $\nu=-2$) with momentum $\bk$ in the mBZ and valley $\tau=\pm$; $N$ is a normalization factor counting the number of unit cells.
We also consider charge density modulations with wavevector $\bq$, $\rho_{\bq} = \frac{1}{N} \sum_{\bk} \langle \psi^\dagger_{\bk + \bq} \tau_0 \psi_{\bk} \rangle$. 
Note that $\rho_{-\bq} = \rho_{\bq}^*$ and $\bI_{-\bq} = \bI_{\bq}^*$.

Under the relevant symmetries of the problem---namely, (spinless) time-reversal $\cT$, $\pi$ rotations $C_{2z}$ around an out-of-plane axis, $U(1)_V$ valley rotations and translations $T_{\bR}$ by moir\'e lattice vectors $\bR$---the fermion operators transform as
\begin{align}
    \cT \psi_{\bk} \cT^{-1} &= \cK \tau_x \psi_{-\bk} ~ ,
    &
    \cC_{2z} \psi_{\bk} \cC_{2z}^{-1} &= \tau_x \psi_{-\bk} \\
    U(1)_V \psi_{\bk} U(1)_V^{-1} &= e^{i \phi \tau_z/2} \psi_{\bk} ~ , 
    &
    T_{\bR} \psi_{\bk} T_{\bR}^{-1} &= e^{i \bk \cdot \bR} \psi_{\bk}.
\end{align}
In the IKS state translation and $U(1)_V$ symmetries are both spontaneously broken; however, the product $T'_{\bR} = T_{\bR} U(1)_V$ is preserved, for a judiciously chosen valley phase rotation that cancels out the contribution from translation, as explained more precisely below.

Under the symmetries outlined above, the IKS order parameters transforms as
\begin{align}
    \cT \bI_q \cT^{-1} &= \bI_q^* ~ , 
    &
    \cC_{2z}\bI_{\bq} \cC_{2z}^{-1} &= \begin{pmatrix}
    I_{\bq}^x \\
    -I_{\bq}^y
    \end{pmatrix}^{\!\!*} \\\label{eqn:IKS_U(1)_sym}
    U(1)_V \bI_{\bq} U(1)_V^{-1} &= R(\phi) \bI_{\bq} ~ , 
    &
    T_{\bR} \bI_{\bq} T_{\bR}^{-1} &= e^{i \bq \cdot \bR} \bI_{\bq}
\end{align}
with $R(\phi$) the usual two-dimensional rotation matrix. In an equivalent formulation with complex order parameters $\Phi^\pm_{\bq} \equiv I_{\bq}^x \pm i I_{\bq}^y$, the transformation properties under valley U(1) rotations and translations amount to multiplication by a phase factor, $e^{i \phi}$ and $e^{i \bq \cdot \bR}$, respectively. In this language it is immediately clear that the IKS order respects the effective translation symmetry $T_{\bR}'$ with the phase chosen as $\phi = - \bq \cdot \bR$.
 The charge density transforms as
\begin{align}
    \cT \rho_{\bq} \cT^{-1} &= \rho_{\bq}^* ~ , 
    &
    \cC_{2z}\rho_{\bq} \cC_{2z}^{-1} &= \rho_{\bq}^* \\
    U(1)_V \rho_{\bq} U(1)_V^{-1} &= \rho_{\bq} ~ ,
    &
    T_{\bR} \rho_{\bq} T_{\bR}^{-1} &= e^{i \bq \cdot \bR} \rho_{\bq}.
\end{align}

Considering only the IKS order parameter and the charge density separately for the moment, the allowed couplings in a Ginzburg-Landau free energy are
\begin{align}
    F^{\rm IKS}_{\bq} =& r_I |\bI_{\bq}|^2 + U_I |\bI_{\bq}|^4  + i \lambda_I {\bf \hat{z}} \cdot(\bI_{\bq} \times \bI_{\bq}^*) %
    + r_{\rho} |\rho_{\bq}|^2 + U_{\rho} |\rho_{\bq}|^4 + \cdots
\end{align}
where all couplings are real (and in general $\bq$ dependent) and we neglected anisotropy terms to fourth order. Note that the $\lambda_I$ term, like all other pieces, manifestly preserves all symmetries including $U(1)_V$.
The leading-order coupling between the charge and IKS modulations is of the form
\begin{equation}\label{eqn:g1_term}
    g_1  \left( (\bI_{\bq} \cdot \bI_{\bq}) \rho_{2 \bq}^* + (\bI_{\bq} \cdot \bI_{\bq})^* \rho_{2 \bq} \right),
\end{equation}
which couples IKS modulations at wavevector $\bq$ with charge modulations at wavevector $2\bq$.  

The free energy captured above varies smoothly with $\bq$, and thus at this level lock-in to commensurate wavevectors will not arise.  We therefore consider another avenue, rooted in the fact that $U(1)_V$ valley rotation symmetry in TTG is only emergent at low energies. Indeed, crystal momentum conservation on the microscopic graphene scale only enforces a $Z_3$ symmetry. If $U(1)_V$ is broken down to $Z_3$, the first expression in Eq.~\eqref{eqn:IKS_U(1)_sym} should be replaced by the weaker condition
\begin{equation}
    Z_3 \bI_{\bq} Z_3^{-1} = R \!\left(\frac{2\pi j}{3} \right) \bI_{\bq}
\end{equation}
with $j=0,1,2$. As a consequence, \emph{for particular commensurate IKS wavevectors} $\bq_{\rm comm}$, a new family of terms of the form
\begin{align}
    \delta_n \left[\sum_{j = 1}^3 ({\bf \hat e}_j\cdot \bI_{\bq_{\rm comm}})^n + c.c. \right] 
\end{align}
with integers $n$ can arise.  Here ${\bf \hat e}_1 = {\bf \hat{x}}$ and ${\bf \hat e}_{2,3}$ are unit vectors in the $(x,y)$ plane rotated by $\pm 2\pi/3$ compared to ${\bf \hat e}_1$---thus ensuring preservation of $Z_3$ symmetry as well as $\mathcal{C}_{2z}$.  Compatibility with $\mathcal{T}$ is guaranteed by the addition of the complex conjugate factors.  Preservation of translation symmetry for a given $\delta_n$, however, requires wavevectors $\bq_{\rm comm}$ that satisfy
$n \bq_{\rm comm} \cdot {\bR} = 2 \pi m$ with $m$ an integer.  Under certain circumstances that we quantify below, IKS states characterized by such commensurate wavevectors can uniquely gain energy from the corresponding $\delta_n$ term---leading to commensurate lock-in.

It is useful to now explicitly examine the structure of the $\delta_n$ terms for different values of $n$.  The $\delta_1$ term vanishes trivially since $\sum_j {\bf \hat e}_j = 0$.  The $\delta_2$ term is proportional to $(\bI_{\bq_{\rm comm}} \cdot \bI_{\bq_{\rm comm}})$, which is symmetry-allowed provided $\bq_{\rm comm} = m\bg/2$ with $\bg$ a (moir\'{e}) reciprocal lattice vector\footnote{Notice that $\delta_2$ has an identical structure to the $g_1$ term in Eq.~\eqref{eqn:g1_term} above---but without the $\rho_{2\bq}$ factors, which are no longer symmetry required.}.
In this case, we have $\bI_{\bq_{\rm comm}} = \bI_{-\bq_{\rm comm}} = \bI_{\bq_{\rm comm}}^*$, allowing us to write  $(\bI_{\bq_{\rm comm}} \cdot \bI_{\bq_{\rm comm}}) = | \bI_{\bq_{\rm comm}}|^2$; thus $\delta_2$ is identical to the $r_I$ term already included in the free energy above and correspondingly does \emph{not} generate commensurate lock-in.  Similar conclusions hold for $\delta_4$, which is proportional to $|\bI_{\bq_{\rm comm}}|^4$ and identical to the $U_I$ term already included above.

In contrast, $\delta_{3,5,6}$ (along with higher $n$'s) generate fundamentally new terms that \emph{do} produce lock-in to commensurate wavevectors $\bq_{\rm comm} = m\bg/3$, $m\bg/5$, and $m\bg/6$, respectively.  To understand how lock-in arises, suppose first that quadratic terms in $F_{\bq}^{\rm IKS}$ favor condensation at some incommensurate wavevector $\bq_0$ that is `nearby' to one of the preceding commensurate wavevectors.  
Lock-in occurs---without fine-tuning---when the free-energy gain from the $Z_3$-symmetric anisotropies accrued from condensing at $\bq_{\rm comm}$ overwhelms the energy cost from $F_{\bq}^{\rm IKS}$ that results from not selecting the otherwise optimal $\bq_0$ wavevector.

While the above mechanism provides a proof-of-concept scenario for commensurate lock-in, estimates of the free energy parameters, including the relevant $\delta_{n}$ terms, are important for assessing resilience of the effect to thermal fluctuations and other experimentally relevant perturbations. We leave such an analysis for future work. In general, we expect that any lock-in energy scale would be larger for wavevectors $\frac{\bg}{3}$ and $\frac{2\bg}{3}$, as they can take advantage of the cubic term in the free energy, as compared to $\frac{\bg}{2}$ which can only gain from the sextic term.

\end{document}